\documentclass[%        Class options:
aps,%                   American Physical Society
prd,%                   Physical Review D
showpacs,%              Displays PACS after abstract
preprint,%              Preprint layout
tightenlines,%          Single spaced lines
superscriptaddress,%   Authors' addresses linked with superscripts
nofootinbib,%           Does not treat footnotes as references
floatfix]%              Fixes float errors
{revtex4}%              REVTEX 4 Package used
\usepackage{graphicx,%  Default Latex 2eps package for embedding figures
longtable%             Useful for long table
}
\begin{document}
%
%===================================================================================
\title{     Status of atmospheric $\nu_\mu\to\nu_\tau$ oscillations and decoherence\\
                        after the first K2K spectral data}
%===================================================================================
%
\author{        G.L.~Fogli}
\affiliation{   Dipartimento di Fisica
                and Sezione INFN di Bari\\
                Via Amendola 173, 70126 Bari, Italy\\}
\author{        E.~Lisi}
\affiliation{   Dipartimento di Fisica
                and Sezione INFN di Bari\\
                Via Amendola 173, 70126 Bari, Italy\\}
\author{        A.~Marrone}
\affiliation{   Dipartimento di Fisica
                and Sezione INFN di Bari\\
                Via Amendola 173, 70126 Bari, Italy\\}
\author{        D.~Montanino}
\affiliation{   Dipartimento di Scienza dei Materiali
                and Sezione INFN di Lecce\\
                Via Arnesano, 73100 Lecce, Italy\\}

\begin{abstract}%...........................................................
We review the status of $\nu_\mu\to\nu_\tau$ flavor transitions of
atmospheric neutrinos in the 92 kton-year data sample collected in
the first phase of the Super-Kamiokande (SK) experiment, in
combination with the recent spectral data from the KEK-to-Kamioka
(K2K) accelerator  experiment (including 29 single-ring muon
events). We consider a theoretical framework which embeds flavor
oscillations plus hypothetical decoherence effects, and where both
standard oscillations and pure decoherence represent limiting
cases. It is found that standard oscillations provide the best
description of the SK+K2K data, and that the associated
mass-mixing parameters are determined at $\pm1\sigma$ (and
$N_\mathrm{DF}=1$)  as: $\Delta m^2=(2.6\pm 0.4)\times 10^{-3}$
eV$^2$ and $\sin^2 2\theta=1.00^{+0.00}_{-0.05}$. As compared with
standard oscillations, the case of pure decoherence is disfavored,
although it cannot be ruled out yet. In the general case,
additional decoherence effects in the $\nu_\mu\to\nu_\tau$ channel
do not improve the fit to the Sk and K2K data, and upper bounds
can be placed on the associated decoherence parameter. Such
indications, presently dominated by SK, could be strengthened by
further K2K data, provided that the current spectral features are
confirmed with higher statistics. A detailed description of the
statistical analysis of SK and K2K data is also given, using the
so-called ``pull'' approach to systematic uncertainties.
\end{abstract}%.............................................................
\medskip
\pacs{%         PACS Numbers:
14.60.Pq, 13.15.+g, 03.65.Yz, 95.55.Vj}
\maketitle

%%%%%%%%%%%%%%%%%%%%%%%%%%%%%%%%%%%%%%%%%%%%%%%%%%%%%%%%%%%%%%%%%%%%%%%%%%%%%%%%
\section{Introduction}
%%%%%%%%%%%%%%%%%%%%%%%%%%%%%%%%%%%%%%%%%%%%%%%%%%%%%%%%%%%%%%%%%%%%%%%%%%%%%%%%

In its first phase of operation (years 1996--2001), the
Super-Kamiokande (SK) experiment has provided, among other
important results, compelling evidence for atmospheric $\nu_\mu$
disappearance \cite{Evid,KaTo}. This evidence, now firmly based on
a high-statistics 92 kton-year exposure \cite{Sh02}, has not only
been corroborated \cite{Ke02} by consistent indications in the
MACRO \cite{MACR} and Soudan~2 \cite{Soud} atmospheric neutrino
experiments, but has also been independently checked by the first
long-baseline KEK-to-Kamioka (K2K) accelerator experiment
\cite{K2K1,K2K2}, using SK as a target for $\nu_\mu$'s produced
250 km away with $\langle E_\nu\rangle \sim 1.3$~GeV.

Neutrino flavor oscillations, originated by nonzero mass-mixing
parameters ($\Delta m^2,\sin^2 2\theta$) in the
$\nu_\mu\to\nu_\tau$ channel, provide by far the best and most
natural (and probably unique) explanation for the observed
$\nu_\mu$ disappearance \cite{Evid,KaTo}. Among the ``exotic''
scenarios which have been proposed as radical alternatives (see
\cite{KaTo,PaVa,Pa00,Ke02} for reviews), at present only
hypothetical neutrino decoherence \cite{Li00} (see also
\cite{Be01,Ga01}) appears as a possible ``survivor'' in the
official SK data analysis \cite{Ke02,Ke03,Sm02}. All the other
models envisaged so far to challenge the standard picture are
strongly rejected as dominant explanations, and are bound to give
at most subdominant effects (see, e.g., \cite{KaTo,PaVa,Ke02}.

In this paper, we review the phenomenological status of both
standard oscillations and decoherence effects in the
$\nu_\mu\to\nu_\tau$ channel, in the light of the latest SK
atmospheric zenith distributions of leptons \cite{Sh02,Ke02,Wi02}
and of the first spectral results from the K2K experiment
\cite{K2K2,Wi02}. In Sec.~II we briefly review the adopted
theoretical framework and describe the data set used. In Sec.~III
and IV we show, respectively, the results of our analysis for the
cases of standard oscillations and pure decoherence, while in
Sec.~V we consider the most general case
(oscillation+decoherence). Conclusions are given in Sec.~VI.

We find that standard oscillations provide the best fit to the
SK+K2K data, and that the mass-mixing parameters  are determined
at $\pm1\sigma$ as
%---------------------------------------------------------------------
\begin{eqnarray}
\label{1st}\Delta m^2/\mathrm{eV}^2&=&(2.6\pm 0.4)\times 10^{-3} ,\\
\label{2nd}\sin^2 2\theta&=&1.00^{+0.00}_{-0.05}\ ,
\end{eqnarray}
%---------------------------------------------------------------------
with errors scaling linearly, with good accuracy, up to
$\sim\!3\sigma$. Conversely, possible additional decoherence
effects are generally disfavored, although not very strongly. We
discuss how the upper bounds on such effects, currently dominated
by SK, can potentially be improved by further K2K spectral data.
The technical details of our SK and K2K data analysis are finally
described in two appendices,  using the so-called ``pull''
approach to systematic uncertainties.

%%%%%%%%%%%%%%%%%%%%%%%%%%%%%%%%%%%%%%%%%%%%%%%%%%%%%%%%%%%%%%%%%%%%%%%
\section{Theoretical framework and experimental data}

Oscillating neutrino systems  may be thought as extremely
sensitive interferometers with long arm lengths. When the
interferometer beams interact with a background (e.g., matter),
the interference pattern may be altered, and may even disappear if
the background is ``fuzzy.'' In theories of quantum gravity, it
has long been speculated that the space-time itself may act as a
``fuzzy'' or ``foamy'' background for any propagating particle,
leading to possible decoherence and dissipative phenomena on
macroscopic scales (see, e.g., the bibliography in \cite{Li00}.)
In this case, the neutrino evolution has to be described in the
language of quantum open systems interacting with a generic
environment \cite{Be01,Ga01}.

In Ref.~\cite{Li00} the formalism of quantum open systems has been
applied to the case of two-family $\nu_\mu\to\nu_\tau$ transitions
of atmospheric neutrinos. Assuming entropy increase and energy
conservation in the neutrino subsystem, the evolution equation was
shown to depend on only one additional (decoherence) parameter,
here denoted as $\mu^2$, in addition to the usual mass-mixing
parameters $(\Delta m^2,\sin^22\theta)$.%
%----------------------------------------------------------------
\footnote{The relation with the parameter $\beta$ in \cite{Li00}
is $\mu^2=2\beta$.}
%-----------------------------------------------------------------
 With the further
assumption of Lorentz-invariance of $\mu^2$, it was shown
\cite{Li00} that  the $\nu_\mu$ survival probability takes, in
natural units, the form
%.....................................................................................
\begin{equation}
\label{Ptot} P_{\mu\mu}(\Delta m^2,\theta,\mu^2)=1-\frac{\sin^2
2\theta}{2}\left[ 1-e^{-\frac{\mu^2
L}{2E_\nu}}\cos\left(\frac{\Delta m^2 L }{2E_\nu}\right)\right]\ ,
\end{equation}
%.....................................................................................
where $E_\nu$ ($L$) is the neutrino energy (pathlength), and
$\mu^2$ has the dimensions of a squared mass. In the absence of
first-principle calculations, $\mu^2$ must be considered as a
purely phenomenological parameter, to be constrained by
experimental data.

The above equation has two interesting limits. The first is
reached for $\mu^2=0$, corresponding to the well-known case of
standard oscillations,
%.....................................................................................
\begin{equation}
\label{Posc} P^\mathrm{osc}_{\mu\mu}(\Delta
m^2,\theta)=1-\frac{\sin^2 2\theta}{2}\left[
1-\cos\left(\frac{\Delta m^2 L }{2E_\nu}\right)\right]\ ,
\end{equation}
%.....................................................................................
which perfectly fits atmospheric $\nu$ data with $\Delta m^2\sim
3\times 10^{-3}$ eV$^2$ and $\sin^2 2\theta\sim 1$. The second
limit is reached for $\mu^2=0$, corresponding to the case of pure
decoherence
%.....................................................................................
\begin{equation}
\label{Pdec} P^\mathrm{dec}_{\mu\mu}(\mu^2,\theta)=1-\frac{\sin^2
2\theta}{2}\left[ 1-\exp\left({-\frac{\mu^2
L}{2E_\nu}}\right)\right]\ ,
\end{equation}
%.....................................................................................
where the oscillation pattern is completely absent and the
$\nu_\mu$ disappearance rate becomes monotonic in $L/E_\nu$.
Surprisingly, the pure decoherence case turned out to fit well the
52 kton-year data SK atmospheric data for $\mu^2 \sim 3\times
10^{-3}$ eV$^2$ and $\sin^2 2\theta\sim 1$, with a $\chi^2$
comparable to the standard oscillation case \cite{Li00}. Only with
almost doubled SK statistics (92 kton-year exposure \cite{Sh02})
some weak indications have started to emerge against the pure
decoherence scenario \cite{Ke02,Ke03,Sm02}.

It is important to note that neutrino decay  \cite{Ba99} can lead
to an exponentially decreasing $P_{\mu\mu}$ qualitatively similar
to Eq.~(\ref{Pdec}). The decay and decoherence scenarios cannot
thus be easily distinguished in the $\nu_\mu$ disappearance
channel. However, they can be distinguished through the appearance
mode, i.e., through neutral current (NC) events and $\tau$
appearance events in SK \cite{Sh02}. In fact, while in the
decoherence case the total number of active neutrinos is
conserved, in the decay scenario it decreases with $L/E_\nu$. The
nonobservation of NC event suppression and the indications for
$\tau$ appearance  in SK can thus be used to reject decay
\cite{Sh02,Ke02,Sm02} but not decoherence effects.

We remark that, in the general case [Eq.~(\ref{Ptot})],
decoherence effects basically acts as a damping factor for the
oscillating term \cite{Li00,Oh01}. Testing atmospheric and
long-baseline accelerator neutrino data through Eq.~(\ref{Ptot})
amounts thus to test how well the oscillatory pattern is favored
by (or hidden in) the data. Therefore, the decoherence scenario
represents a useful benchmark, although its theoretical
motivations are admittedly weaker than those supporting standard
oscillations. In practice, in fits with unconstrained $(\Delta
m^2,\mu^2,\sin^22\theta)$, the emergence of the standard
oscillatory pattern should be signalled by a preference for small
values of $\mu^2$. As we shall see, SK and K2K consistently favor
$\mu^2\simeq 0$; however, the absence of a clear oscillation
pattern in the data makes this indication not very strong yet.

We conclude this section with a review of the SK and K2K data sets
used for our analysis (see the appendices for more details).
Concerning SK atmospheric neutrino data (92 kton-year
\cite{Sh02,Ke02,Wi02}), we use the following zenith angle
$(\theta_z)$ distributions of leptons: sub-GeV $e$-like and
$\mu$-like events (SG$e$ and SG$\mu$), divided in 10+10 bins;
multi-GeV $e$-like and $\mu$-like events (MG$e$ and MG$\mu$),
divided in 10+10 bins; upward stopping and through-going $\mu$
events (US$\mu$ and UT$\mu$), divided in 5+10 bins. The
calculation of the theoretical events rates $R_n^\mathrm{theo}$ in
each of the 55 bins is done as in \cite{Fo98,Fo01,Ma01}. The SK
statistical analysis is considerably improved with respect to
\cite{Fo98,Ma01}, where only three sources of systematic errors
were used (leading to correlated uncertainties in energy, angle,
and flavor mis-identification \cite{Fo98}). Now the set of
systematic errors has been enlarged to 11 entries, leading to a
more complex structure of correlated errors affecting the
$R_n^\mathrm{theo}$'s, with allowance for various kinds of
(relative) normalization and shape uncertainties. As emphasized in
\cite{Gett} for the solar neutrino case, systematic uncertainties
can be implemented in the $\chi^2$ statistics through two
equivalent methods: 1) by building the total covariance matrix
(``covariance method''), or 2) by adding quadratic penalties in
the systematic pulls (``pull method''). The latter approach
(adopted in this work) allows to study how systematic errors alter
the theoretical predictions from the central values
$R_n^\mathrm{theo}$ to ``shifted'' values $\overline
R_n^\mathrm{theo}$, in order to match the data. The difference
$\overline R_n^\mathrm{theo}-R_n^\mathrm{theo}$ is thus useful to
gauge the size and the direction of systematic effects in the data
fit. More precise definitions and technical details are given in
Appendix~A.

Concerning the K2K data, we use the absolute spectrum of muon
events in terms of the reconstructed neutrino energy $E$
\cite{K2K2,Wi02}, which provides a total of 29 events (here
divided in 6 bins). In this sample, the parent neutrino
interactions are dominantly quasi-elastic (QE), and the
reconstructed energy $E$ is thus closely correlated with the true
neutrino energy $E_\nu$.
%----------------------------------------------------------------
\footnote{We cannot use the full K2K data sample (56 events
\cite{K2K2}), which contains 27 additional events of dominantly
non-QE origin, whose analysis is basically not reproducible
outside the K2K collaboration.}
%---------------------------------
 In each bin, we attach to the theoretical
prediction $N_n^\mathrm{theo}$ the two leading systematic errors
(due to near-far and normalization uncertainties) through the pull
approach. As for the SK analysis, we also discuss the
systematically shifted theoretical predictions $\overline
N_n^\mathrm{theo}$. The relevant technical details are given in
Appendix~B.

%%%%%%%%%%%%%%%%%%%%%%%%%%%%%%%%%%%%%%%%%%%%%%%%%%%%%%%%%%%%%%%%%%%%%%%
\section{Standard Oscillations}
%%%%%%%%%%%%%%%%%%%%%%%%%%%%%%%%%%%%%%%%%%%%%%%%%%%%%%%%%%%%%%%%%%%%%%%%%

In this section we discuss updated bounds on the parameters
$(\Delta m^2,\,\sin^2 2\theta)$, governing the scenario of
standard oscillations. The most important result is a significant
strengthening of the upper bound on $\Delta m^2$ induced by K2K
data.

\begin{figure}
\vspace*{-2.5cm}\hspace*{-1.8cm}
\includegraphics[scale=0.92, bb= 30 100 500 700]{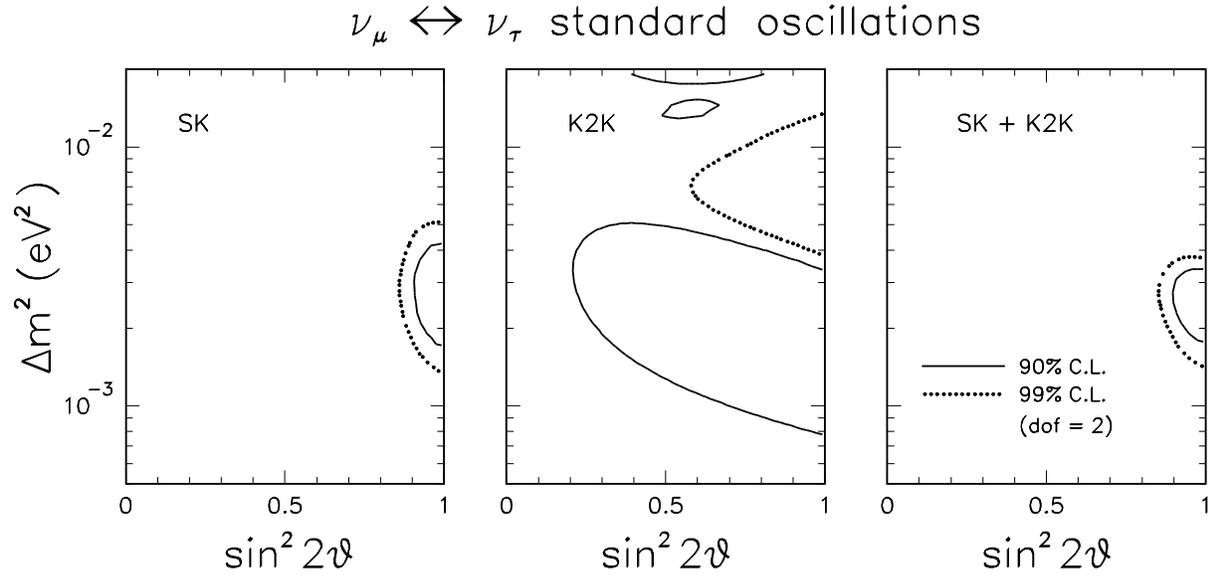}
\vspace*{-2cm} \caption{\label{fig01} Standard oscillations in the
$\nu_\mu\to\nu_\tau$ channel: Bounds on the parameters $(\Delta
m^2,\,\sin^2 2\theta)$ from SK atmospheric data (left panel), K2K
spectral data (middle panel), and their combination (right panel).
The solid and dotted curves refer, respectively, to the 90\% and
99\% C.L.\ contours for $N_\mathrm{DF}=2$ ($\Delta\chi^2=4.61$ and
9.21).}
\end{figure}

Figure~\ref{fig01} shows the joint bounds on the $(\Delta
m^2,\,\sin^2 2\theta)$ parameters from our analysis of SK, K2K,
and SK+K2K data, derived through $\Delta\chi^2$ cuts around the
$\chi^2$ minima. The minima and their positions are reported in
the ``standard oscillation'' columns of Table~\ref{chi2val}. The
global SK best fit $(\chi^2_{\min}=34.8)$ might seem somewhat too
good, as compared with the $N_\mathrm{DF}=55-2$ degrees of
freedom. Howewer, this value is only $1.8\sigma$ below the typical
$\chi^2$ expectations ($N_\mathrm{DF}\pm\sqrt{2\,N_\mathrm{DF}}$),
and is thus not suspicious from a statistical viewpoint.

%%%%%%%%%%%%%%%%%%%%%%%%%%%%%%%%%%%%%%%%%%%%%%%%%%%%%%%%%%%%%%%%%%%%%%%%%%%%%%%%%%%%%
\begingroup \squeezetable
\begin{table}[t]
\caption{\label{chi2val} \footnotesize\baselineskip=6pt
Coordinates and values of the absolute $\chi^2$ minima for
separate and combined fits to SK and K2K data, in the scenarios of
standard oscillation, pure decoherence, and oscillations plus
decoherence.}
\newcommand{\T}{$\times 10^{-3}$}
\begin{ruledtabular}
\begin{tabular}{c|ccc|ccc|ccc}
& \multicolumn{3}{c|}{Standard oscillations}&
\multicolumn{3}{c|}{Pure decoherence} &
\multicolumn{3}{c}{Oscillations plus decoherence}\\
Best fits & SK & K2K & SK+K2K & SK & K2K & SK+K2K &  SK & K2K &
SK+K2K\\
\hline %
$\Delta m^2$~(eV$^2$)&2.72\T& 2.65\T&2.63\T& ---  & ---   & ---  &2.72\T&2.65\T&2.63\T \\ %
$\sin^2 2\theta$     &1.00  & 0.82  &1.00  & 1.00 & 0.86  & 1.00 &1.00  &0.82  &1.00   \\ %
$\mu^2$~(eV$^2$)     & ---  & ---   & ---  &2.42\T& 2.46\T&2.44\T& 0.00 & 0.00 & 0.00  \\ %
$\chi^2$             &34.8  & 9.4   &45.2  &42.1  & 11.5  &54.1  & 34.8 & 9.4  & 45.2  \\
\end{tabular}
\end{ruledtabular}
\end{table}
\endgroup
%%%%%%%%%%%%%%%%%%%%%%%%%%%%%%%%%%%%%%%%%%%%%%%%%%%%%%%%%%%%%%%%%%%%%%%%%%%%%%%%%%%%%%%%
In Fig.~\ref{fig01}, the bounds in the left panel are very close
to the official SK ones, as presented in \cite{Sh02,Ke02,Wi02}.
The bounds in the middle panel are instead slightly weaker than
the official K2K ones \cite{K2K2}, especially in terms of $\sin^2
2\theta$. In particular, we do not find a lower bound on
$\sin^22\theta$ at 99\% C.L.\ (for $N_\mathrm{DF}=2$). The reason
is that we cannot use the additional (dominantly) non-QE event
sample of K2K (27 events), which would help to constrain the
overall rate normalization and thus $\sin^2 2\theta$. This fact
might also explain why we find the K2K best
fit at $\sin^2 2\theta=0.82$ rather than at 1.00 as in \cite{K2K2}.%
%-------------------------------------------------------------
\footnote{In any case, this difference is not statistically
significant, since the $\chi^2$ increase from our K2K best-fit
point to maximal mixing is $\lesssim 1$.}
%--------------------------------------------------------------
Due to the slight anticorrelation between $\Delta m^2$ and $\sin^2
2\theta$ in the K2K analysis (middle panel of Fig.~\ref{fig01}),
an increase of $\sin^2 2\theta$ from 0.82 to 1.00 generates a
slight decrease in the K2K favored range for $\Delta m^2$, which
is reflected in the SK+K2K best-fit $\Delta m^2$ (see
Table~\ref{chi2val}).

By comparing the left and right panels of Fig.~\ref{fig01}, the
main effect of K2K appears to be the strengthening of the upper
bound on $\Delta m^2$, consistently with the trend of the first
K2K data (rate only \cite{K2K1}, no spectrum) \cite{Ma01}. The
main reason is that, for $\Delta m^2\sim (4$--$6)\times 10^{-3}$
eV$^2$, the first oscillation minimum would be located at---or
just above---the K2K energy spectrum peak, implying a strong local
and overall suppression of the expected events, contrary to the
K2K observations.

\begin{figure}
\vspace*{+2cm}\hspace*{-2.6cm}
\includegraphics[scale=0.92, bb= 30 100 500 700]{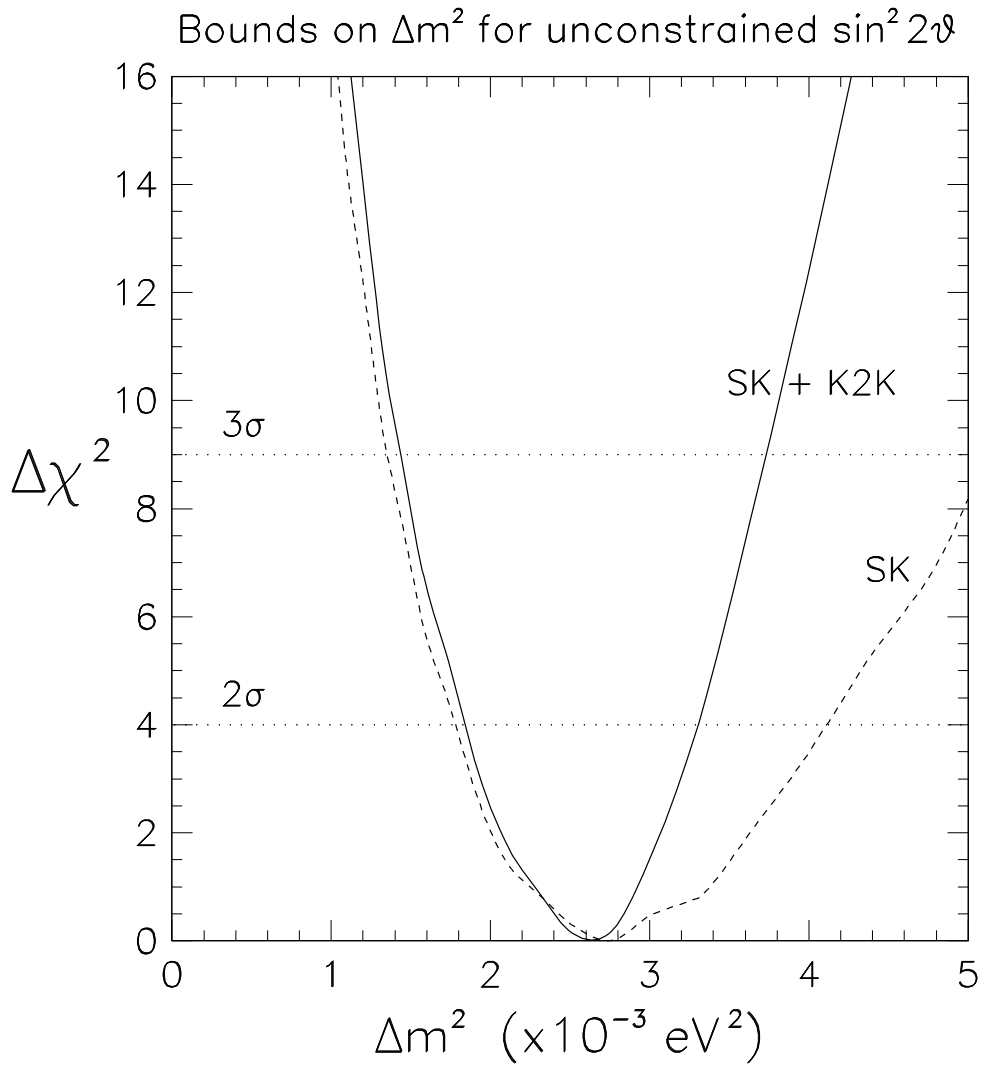}
\vspace*{-3cm} \caption{\label{fig02} Standard oscillations in the
$\nu_\mu\to\nu_\tau$ channel: Bounds on $\Delta m^2$ for
unconstrained $\sin^22\theta$ from SK (dashed curve) and SK+K2K
(solid curve). The intersections with the horizontal dotted lines
give the $2\sigma$ and $3\sigma$ (upper and lower) bounds on
$\Delta m^2$ for $N_\mathrm{DF}=1$. By fitting the SK+K2K curve
with a parabola, the $\pm 1\sigma$ interval is derived as $\Delta
m^2=(2.6 \pm 0.4)\times 10^{-3}$ eV$^2$.}
\end{figure}

Figure~\ref{fig02} shows the SK and SK+K2K bounds on $\Delta m^2$,
when the $\sin^2 2\theta$ parameter is projected (minimized) away.
The linear scale in $\Delta m^2$ makes the K2K impact on the upper
limit more evident. Notice that, up to $\sim\!3\sigma$, the global
(SK+K2K)  $\chi^2$ function is approximately parabolic in the {\em
linear\/} variable $\Delta m^2$, so that one can properly defined
a one-standard-deviation error for this parameter. This useful
feature of the SK+K2K fit was already argued on the basis of a
graphical reduction of the official SK and K2K likelihood
functions \cite{Le02}, and is here confirmed through a full
analysis. By keeping only the first significant figure in the
error estimate, a parabolic fit provides the $\pm 1\sigma$ range
%---------------------------------------------------------------
\begin{equation}
\label{Dm2range}
 \Delta m^2 = (2.6 \pm 0.4)\times 10^{-3}\mathrm{\ eV}^2\ ,
\end{equation}
%----------------------------------------------------------------
with $\pm N\sigma$ errors scaling linearly with $N$ (up to $N\simeq 3$).%
%-----------------------------------------------------------------
\footnote{This range is consistent with the estimate $\Delta
m^2=(2.7\pm 0.4)\times 10^{-3}$ eV$^2$ of \cite{Le02}, derived
from a less detailed analysis.}
%-----------------------------------------------------------------

\begin{figure}
\vspace*{+2cm}\hspace*{-2.6cm}
\includegraphics[scale=0.92, bb= 30 100 500 700]{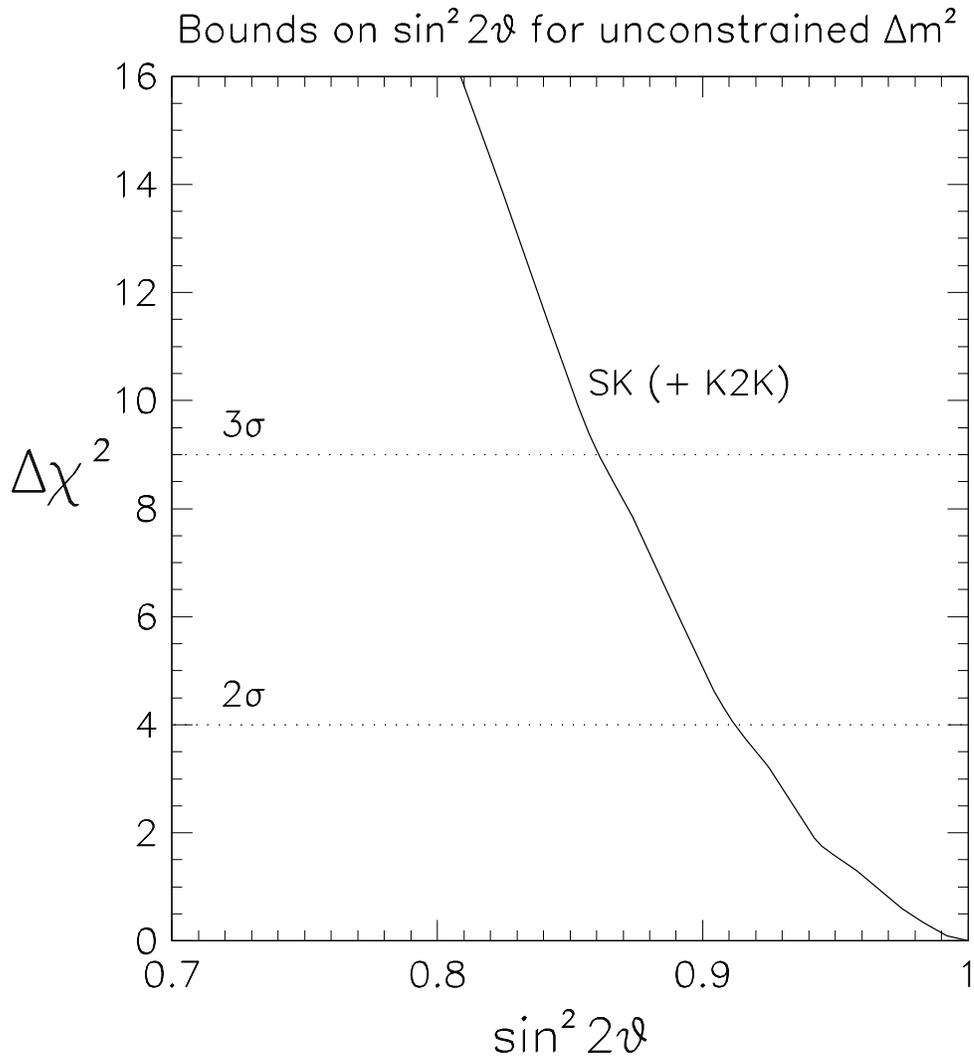}
\vspace*{-3cm} \caption{\label{fig03} Standard oscillations in the
$\nu_\mu\to\nu_\tau$ channel: Bounds on $\sin^2 2\theta$ for
unconstrained $\Delta m^2$ from SK data. The inclusion of K2K data
induces negligible changes (not shown). The intersections with the
horizontal dotted lines give the $2\sigma$ and $3\sigma$ lower
bounds on $\sin^2 2\theta$ for $N_\mathrm{DF}=1$. By fitting the
curve with a parabola, the $\pm 1\sigma$ interval is derived as
$\sin^2 2\theta=1.00^{+0.00}_{-0.05}$, where only the first
significant digit is kept in the lower error.}
\end{figure}

The bounds on $\sin^2 2\theta$ are instead entirely dominated by
SK. Figure~\ref{fig03} shows the  $\Delta\chi^2$ function in terms
of $\sin^2 2\theta$, for $\Delta m^2$ projected (minimized) away
in the SK fit. In this figure, the addition of K2K data would
insignificantly change the bounds (not shown), which thus hold for
both the SK and the SK+K2K fit. Also in this case, the nearly
parabolic behavior of $\Delta \chi^2$ allows to properly define a
$1\sigma$ range,
%---------------------------------------------------------------
\begin{equation}
\label{sinrange} \sin^2 2\theta = 1.00^{+0.00}_{-0.05} \ ,
\end{equation}
%----------------------------------------------------------------
with the lower $N\sigma$ error scaling linearly with $N$ (up to
$N\simeq 3$).%
%---------------------------------------------------
\footnote{The ``upper'' (null) error is, of course, trivial.}
%----------------------------------------------------
Equations~(\ref{Dm2range}) and (\ref{sinrange}) concisely review
the current fit to the standard oscillation parameters, as
anticipated in the Introduction.

\begin{figure}
\vspace*{-1.5cm}\hspace*{-3cm}
\includegraphics[scale=0.92, bb= 30 100 500 700]{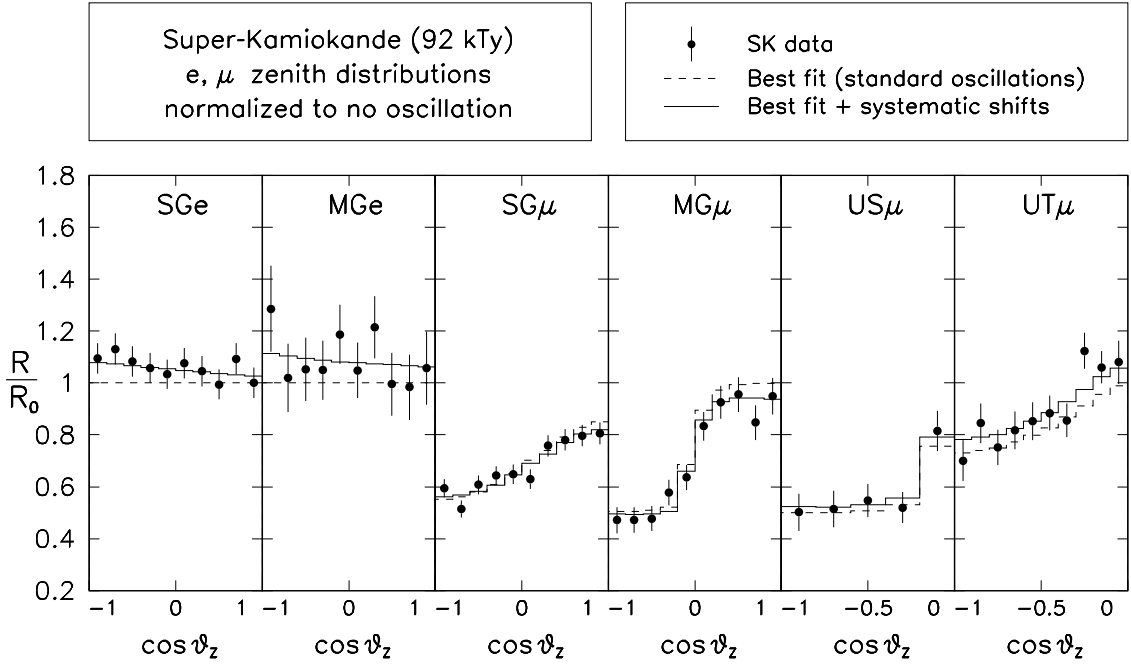}
\vspace*{-3cm} \caption{\label{fig04} Standard oscillations in the
$\nu_\mu\to\nu_\tau$ channel: SK experimental zenith distributions
($\pm 1\sigma_\mathrm{stat}$), compared with the corresponding
theoretical ones at the global (SK+K2K) best-fit point given in
Table~\ref{chi2val}. All distributions are normalized to the
unoscillated predictions in each bin. For the theoretical event
rates, we show both the central values $R_n^\mathrm{theo}$ (dashed
histograms) and the ``shifted'' values $\overline
R_n^\mathrm{theo}$ (solid histograms), which embed the effect of
systematic pulls. The difference between $\overline
R_n^\mathrm{theo}$ and $R_n^\mathrm{theo}$ shows how much (and in
which direction) the correlated systematic errors tend to stretch
the predictions in order to match the data. See the text and
Appendix~A  for details.}
\end{figure}

Figure~\ref{fig04} shows the comparison between observations and
best-fit predictions for the SK zenith distributions. The best fit
point refers to the case of SK+K2K standard oscillations in
Table~\ref{chi2val}. Since the very good agreement between data
and theory is no longer a surprise, in the following we comment on
the ``fine structure'' of the SK data fit. This requires, however,
that the reader can grasp the difference between theoretical
predictions with and without the shifts induced by correlated
systematics (solid and dashed histograms in Fig.~\ref{fig04},
respectively), which is explained in detail in Appendix~A.

In Fig.~\ref{fig04}, the comparison between solid and dashed
histograms shows that the systematic shifts are often comparable
in size to the statistical errors, implying that just increasing
the SK atmospheric $\nu$ statistics will hardly bring decisive new
information on the standard oscillation scenario (or on physics
beyond it). In the SG and MG samples, the fit clearly exploits the
systematic uncertainties to increase the $e$-like event
normalization, especially in the upward direction, so as to
reduce the ``electron excess'' possibly indicated by SK data.%
%-------------------------------------------------------------------
\footnote{This tendency could be exacerbated by adopting the most
recent atmospheric flux calculations, which consistently predict a
lower normalization at relatively low energies
\cite{Ho03,Ga03,Ho01,Ba00}. Nevertheless, preliminary results
indicate that the induced changes in the $(\Delta m^2,\sin^2
2\theta)$ bounds would be minimal \cite{Ka02}.}
%--------------------------------------------------------------------
 Concerning $\mu$-like events in the
SG and MG samples, the fit show an opposite tendency to slightly
decrease the normalization of (especially down-going) events. The
tendency appears to be reversed in the high-energy UT sample.
Taken together, these opposite shifts of $e$-like and $\mu$-like
expectations in the SG and MG samples seem to suggest some
systematic deviation from the predicted $\mu/e$ flavor ratio
which, although not statistically alarming (see the last paragraph
of Appendix~A), should be kept in mind. In fact, deviations of
similar size might have their origin in neutrino physics beyond
$2\nu$ oscillations, e.g.: 1) subleading $3\nu$ oscillations,
which could slightly increase the $e$-like event rates (see, e.g.,
\cite{Fo98,Fo01,Pere}), or 2) subleading (short-baseline)
$\nu_\mu\to\nu_\mu$ oscillations in the so-called (and still alive
\cite{Giun,Stru}) 3+1 scenario, which might slightly alter the
down-going muon rates, as well as the front-detector K2K rate
\cite{Lave,PeSm,Corn}. Unfortunately, since such effects are
typically not larger than the systematic shifts in
Fig.~\ref{fig04}, they are likely (if any) to remain hidden in
higher-statistics SK data, unless a significant reduction of the
current systematics can be accomplished. The happy side of the
story is that, for the same reasons, typical subleading effects
beyond standard $2\nu$ oscillations do not significantly alter the
fit results in Eqs.~(\ref{Dm2range}) and (\ref{sinrange}).

Figure~\ref{fig05} shows the comparison between data and theory
for the K2K absolute spectrum of events, for the same oscillation
best-fit point as in Fig.~\ref{fig04}. In this case, the amount of
systematic deviations preferred by the fit is much smaller than
the current statistical error, implying that there is a great
potential for improvements with higher K2K statistics.

%%%%%%%%%%%%%%%%%%%%%%%%%%%%%%%%%%%%%%%%%%%%%%%%%%%%%%%%%%%%%%%%%%%%%%%
\section{Pure decoherence (and comparison with oscillations)}
%%%%%%%%%%%%%%%%%%%%%%%%%%%%%%%%%%%%%%%%%%%%%%%%%%%%%%%%%%%%%%%%%%%%%%%%%

In this section we discuss the bounds on the parameters
$(\mu^2,\,\sin^2 2\theta)$, governing the scenario of pure
decoherence [Eq.~\ref{Pdec}]. We also discuss how this scenario
compares with standard oscillations in fitting SK and K2K data.

Figure~\ref{fig06} shows the joint bounds on the $(\mu^2,\,\sin^2
2\theta)$ parameters from our analysis of SK, K2K, and SK+K2K
data, derived through $\Delta\chi^2$ cuts around the $\chi^2$
minima, as reported in the ``pure decoherence'' columns of
Table~\ref{chi2val}. Such bounds are clearly dominated by SK. The
best fit values of $\mu^2$ are numerically close to the previous
ones for $\Delta m^2$, since for both oscillations and decoherence
one gets large $\nu_\mu$ disappearance effects  for $xL/E_\nu\sim
O(1)$, where $x=\mu^2$ or $\Delta m^2$ \cite{Li00}.

The global SK+K2K best fit for pure decoherence
$(\chi^2_{\min}=54.1)$ is $\sim\!9$ units higher than for standard
oscillations $(\chi^2_{\min}=45.2)$, where  $\sim\!7$ units are
provided by SK alone.%
%-------------------------------------------------------------
\footnote{The SK collaboration finds a larger difference
($\Delta\chi^2=10.5$ \cite{Ke02,Ke03,Sm02}) by using more data.
Notice that pure decoherence satisfies the goodness-of-fit test
[$\chi^2_\mathrm{min}/N_\mathrm{DF}\sim O(1)$] in both fits (SK
and ours).}
%---------------------------------------------------------------
On the one hand, this difference shows that the data start to have
some sensitivity to decoherence effects, and tend to globally
disfavor them. On the other hand, this sensitivity is not
sufficient to claim rejection of such effects on a purely
phenomenological basis. The features that tend to disfavor
decoherence are better explained in terms of the best-fit
theoretical distributions.

Figure~\ref{fig07} shows the SK zenith distributions for the
global (SK+K2K) best fit to pure decoherence (see
Table~\ref{chi2val}), and should be compared with the analogous
Fig.~\ref{fig04} for the case of standard oscillations. The
differences are hard to detect at first sight. In the decoherence
case, the only noticeable difference  is a slightly less
pronounced suppression of the muon rates for increasing $L$, which
leads to slightly less tilted $\mu$ distributions, as compared
with standard oscillations. Notice also that the pure decoherence
formula (\ref{Pdec}) implies $P^\mathrm{dec}_{\mu\mu}\geq 1/2$
always, while the standard oscillation formula (\ref{Posc}) admits
minima as low as $P^\mathrm{osc}_{\mu\mu}=0$, which can help to
get a more efficient suppression (unless the limit of averaged
oscillations is reached). The small differences in the upgoing
muon distribution slopes are then responsible for the
$\Delta\chi^2\simeq 7$ difference between the two scenarios in SK.
We do not expect that additional NC-enriched data in SK can add
significant contributions to this difference, since (contrary to
the decay scenario \cite{Ba99}) decoherence preserves the number
of active neutrinos, as previously observed. Moreover, given the
size of the systematic shifts in both Figs.~\ref{fig07} and
\ref{fig04}, it seems to us that higher SK statistics will not be
decisive to disentangle decoherence from oscillations. We conclude
that SK tend to disfavor the pure decoherence hypothesis through
its accurate measurements of the upgoing muon distribution shapes;
however, there seems to be little room for significant
improvements with longer SK exposure.

Figure~\ref{fig08} shows the K2K absolute spectrum of events for
the global (SK+K2K) best fit to pure decoherence (see
Table~\ref{chi2val}), to be compared with the analogous
Fig.~\ref{fig05} for standard oscillations.  The differences
between the theoretical predictions in these two Figures are
significant, although the large statistical error bars reduce
their effect to a mere $\Delta\chi^2\simeq 2$ difference between
the oscillation and decoherence fits. The inequality
$P^\mathrm{dec}_{\mu\mu}\geq 1/2$ is crucial to provide a slightly
worse fit in the decoherence case, since it forbids a significant
overall suppression, which can instead be more easily achieved
with oscillations, particularly at low energy. Therefore, there is
room for significant improvements in the discrimination of the two
scenarios with higher K2K statistics,  by looking at the
low-energy part of the spectrum; in fact,  a local rate
suppression $<1/2$ would definitely rule out the decoherence
hypothesis.

Higher K2K statistics will also allow a less coarse binning in the
spectrum analysis. For this reason, we show in Figs.~\ref{fig09}
and \ref{fig10} the unbinned K2K spectrum for representative cases
of standard oscillations and pure decoherence, respectively (at
maximal mixing). The two scenarios show increasing differences for
decreasing energy in all cases. Moreover, within the standard
scenario (Fog.~\ref{fig09}), the spectrum shows rapid shape
variation with $\Delta m^2$. A judicious choice of binning,
especially  in the low-energy part of the spectrum, might thus
enhance the K2K discrimination power in future data analyses, and
might also improve the determination of $\Delta m^2$ in the
standard oscillation case.

%%%%%%%%%%%%%%%%%%%%%%%%%%%%%%%%%%%%%%%%%%%%%%%%%%%%%%%%%%%%%%%%%%%%%%%
\section{OScillation plus decoherence}
%%%%%%%%%%%%%%%%%%%%%%%%%%%%%%%%%%%%%%%%%%%%%%%%%%%%%%%%%%%%%%%%%%%%%%%%%

In this section we consider the general case of oscillations plus
decoherence [Eq.~(\ref{Ptot})], in order to check whether
subdominant (rather than dominant) decoherence effects can help
the fit to the data. It turns out that, by leaving $(\Delta
m^2,\sin^2 2\theta,\mu^2)$  free, the best fit is reached for
$\mu^2=0$ for both SK, K2K, and SK+K2K, as reported in the last
three columns of Table~\ref{chi2val}. The corresponding joint
bounds on the model parameters are shown in Fig.~\ref{fig11},
through projections onto the three coordinate planes.  Standard
oscillations $(\mu^2=0)$ and nearly maximal mixing are clearly
favored. However, the pure decoherence limit ($\Delta m^2=0$) is
still marginally allowed by the data. As discussed in the previous
section, further K2K statistics might help to disfavor this limit
with higher confidence.

Since $\mu^2=0$ is consistently preferred, it makes sense to
derive upper bounds on $\mu^2$, by projecting away the other two
parameters $(\Delta m^2,\sin^22\theta)$ in the fit.
Figure~\ref{fig12} shows the results of this excercise, in terms
of the function $\Delta\chi^2(\mu^2)$. The upper bound at
$3\sigma$ corresponds to $\mu^2\simeq 3\times 10^{-3}$ eV$^2$.
Values of $\mu^2\sim \mathrm{few}\times 10^{-3}$ eV$^2$ are
sufficiently low to be compatible (as remarked in \cite{Li00})
with the non-observation of $\nu_\mu\to\nu_\tau$ appearance in
CHORUS and NOMAD \cite{Zu02}.

We conclude by observing that a really general ``oscillation plus
decoherence'' scenario should be performed with (at least) three
neutrino families, in order to incorporate also solar and reactor
neutrino flavor transitions, and possibly other phenomenological
constraints. Models of this kind would imply several new
decoherence parameters, as well as nontrivial complications
related to unavoidable matter effects. However, given the robust
theoretical and phenomenological basis for standard oscillations,
and the lack of signals {\em in favor\/} of decoherence effects
(so far), a time-consuming analysis of such extended models is
perhaps unwarranted at present. It could become interesting,
however, if further K2K data, plus future long-baseline
accelerator data, would fail to show a clear oscillation pattern.

%%%%%%%%%%%%%%%%%%%%%%%%%%%%%%%%%%%%%%%%%%%%%%%%%%%%%%%%%%%%%%%%%%%%%%%
\section{Summary and conclusions}
%%%%%%%%%%%%%%%%%%%%%%%%%%%%%%%%%%%%%%%%%%%%%%%%%%%%%%%%%%%%%%%%%%%%%%%%%

We have analyzed in detail the current SK atmospheric neutrino
data and and the first K2K spectral data, in order to review the
status of standard $\nu_\mu\to\nu_\tau$ oscillations, as well as
of a possible ``rival'' scenario based on neutrino decoherence. We
have provided updated bounds for the standard oscillation
parameters [Eqs.~(\ref{1st}) and (\ref{2nd})], and have found no
evidence for decoherence effects. However, the SK+K2K data are not
accurate enough to rule out the pure decoherence case yet. The
(currently weak) indications against such effects appear to be
currently dominated by SK data, but the statistical analysis of
the uncertainties reveals that K2K will lead further progress in
this field, especially through higher-statistics tests of the
low-energy spectrum bins. In conclusion, standard
$\nu_\mu\to\nu_\tau$ oscillations are in good health, and
seemingly no alternative scenario is able to provide a better fit
to the SK+K2K data. However, the ``second best'' explanation of
SK+K2K data (i.e., pure decoherence) is still statistically
acceptable, and can provide a useful phenomenological benchmark to
test the emergence of the---still hidden---standard oscillation
pattern.

\acknowledgments

This work was in part supported by the Italian MIUR and INFN
within the ``Astroparticle Physics'' research project. One of us
(E.L.) would like to thank the Organizers of the Triangle meeting
in Vienna, of the Christmas Workshop in Madrid, and of the NOON
Workshop in Kanazawa---where preliminary results from this work
were presented---for kind hospitality.

\appendix
%%%%%%%%%%%%%%%%%%%%%%%%%%%%%%%%%%%%%%%%%%%%%%%%%%%%%%%%%%%%%%%%%%%%%%%%%%
\section{Statistical analysis of SK data}

In this paper, the calculation of the SK zenith distributions of
$e$-like and $\mu$-like event rates is performed as described in
previous works \cite{Fo98,Fo01,Ma01}. In the absence of
oscillations, we normalize our absolute rates to the official 92
kTy SK estimates \cite{Sh02}, which are currently based on the
one-dimensional $\nu$ fluxes by Honda et
al.\ \cite{Ho95} for all classes of events.%
%------------------------------------------------------------------
\footnote{The one-dimensional $\nu$ fluxes by the Bartol group
\cite{Ag96} are used by the SK collaboration for comparison with
the default Honda fluxes.}
%---------------------------------------------------------------------

Our SK statistical analysis, previously based on a simplified
approach to error correlations \cite{Fo98}, has been significantly
improved in two aspects: 1) it has been made more consistent with
the error estimates of the SK collaboration, as reported in
\cite{Kame} (see also \cite{Kiba,Ishi,Mess}); and 2) it has been
cast in a ``pull'' rather than ``covariance'' form, as advocated
in \cite{Gett} for solar neutrinos.

We remind that the ``pull'' approach to correlated systematic
uncertainties (see \cite{Gett} and references therein) amounts to
shift the theoretical rate $R_n^\mathrm{theo}$ in the $n$-th bin
through a set of deviations $\xi_k c^n_k$, where $c^n_k$ is the
$1\sigma$ error associated to the $k$-th source of systematics (in
the present case, $k=1,\dots,11$), and the $\xi_k$'s are random
variables,
%..........................................................................
\begin{equation}
\label{Rtilde} R_n^\mathrm{theo}\to \tilde
R_n^\mathrm{theo}=R_n^\mathrm{theo}+\sum_{k=1}^{11}\xi_k c_n^k\ .
\end{equation}
%.........................................................................

The $\chi^2$ function is then obtained by minimization over the
$\xi_k$'s,
%..........................................................................
\begin{equation}
\label{chiSK}\chi^2_\mathrm{SK}=\min_{\{\xi_k\}} %
\left[ \sum_{n=1}^{55}\left(\frac{\tilde
R_n^\mathrm{theo}-R_n^\mathrm{expt}}{\sigma_n^\mathrm{stat}}\right)^2+
\sum_{k,h=1}^{11} \xi_k \,[\rho^{-1}]_{hk} \,\xi_h \right]\ ,
\end{equation}
%.........................................................................
where $\sigma_n^\mathrm{stat}$ represent the statistical errors
(or, in general, the quadratic sum of all uncorrelated errors) for
the $n$-th bin. Notice that, with respect to the discussion in
\cite{Gett}, we now give allowance for correlations $\rho_{hk}$ of
the
systematic error sources.%
%-----------------------------------------------------------------------
\footnote{In \cite{Gett} all error sources were strictly
independent, so that $\rho_{hk}=\delta_{hk}$. Generalizing to
$\rho_{hk}\neq\delta_{hk}$ leads to the appearance of the inverse
correlation matrix $[\rho^{-1}]_{hk}$ in the last term of
Eq.~(\ref{chiSK}).}
%------------------------------------------------------------------------
 The above $\chi^2$ function is exactly equivalent to build a
quadratic form in the unshifted differences
$(R_n^\mathrm{theo}-R_n^\mathrm{expt})$, with a covariance matrix
defined as
%..........................................................................
\begin{equation}
\label{sigma}\sigma^2_{nm}=
\delta_{nm}\,\sigma_n^\mathrm{stat}\sigma_m^\mathrm{stat}
+\sum_{k,h=1}^{11}\rho_{hk}\,c_n^h\, c_m^k \ .
\end{equation}
%...........................................................................
The minimization in Eq.~(\ref{chiSK}) leads to a solvable set of
linear equations in the $\xi_k$'s, whose solution $\overline\xi_k$
can provide useful information on the role of systematic
uncertainties in the fit \cite{Gett}. For instance, it can be
interesting to study how much the best-fit shifted rates
%..........................................................................
\begin{equation}
\label{Rbar} \overline
R_n^\mathrm{theo}=R_n^\mathrm{theo}+\sum_{k=1}^{11}\overline \xi_k
c_n^k
\end{equation}
%.........................................................................
differ from the standard ones ($R_n^\mathrm{theo}$), as a
consequence of the systematic pulls $\overline\xi_k$.%
%--------------------------------------------------------------------------
\footnote{We observe that the SK collaboration basically shows the
shifted $\overline R_n^\mathrm{theo}$'s in the best-fit plots,
while we have graphically reported the unshifted
$R_n^\mathrm{theo}$'s in previous works \cite{Fo98,Fo01,Ma01}. For
the sake of comparison,  Fig.~\ref{fig04} and \ref{fig07} in this
paper show both representations.}
%-------------------------------------------------------------------------
This valuable information would be lost in the covariance
approach.

Our characterization of the 11 sources of SK uncertainties follows
(or, at least, is inspired by) the detailed error description
given in \cite{Kame} (see also \cite{Kiba,Ishi,Mess}). There is,
however, one important difference, related to the fact that the SK
analysis is based on Monte~Carlo (MC) simulations, while ours is
based on direct calculations of the lepton rates. More precisely,
while in the SK official analysis the systematic uncertainties act
as re-weighting factors for each MC event
\cite{Kame,Kiba,Ishi,Mess}, in our case they act directly on the
observable rates $R_n$ ($n=1,\dots,55$), and are thus completely
characterized by the $55\times 11$ matrix $c_n^k$.

%==================================================================================
\begingroup \squeezetable
\begin{table}[t]
\newcommand{\m}{$-$}
\newcommand{\p}{$+$}
\newcommand{\s}{\phantom{$+$}}
\newcommand{\sE}{SG$e$}
\newcommand{\sM}{SG$\mu$}
\newcommand{\mE}{MG$e$}
\newcommand{\mM}{MG$\mu$}
\newcommand{\pM}{US$\mu$}
\newcommand{\tM}{UT$\mu$}
\setlength\LTleft{0pt}
\setlength\LTright{0pt}\setlength\LTcapwidth{468pt}
\begin{longtable}{@{\extracolsep{10.4pt}}cccrrrrrrrrrrr}
\caption{\label{SKsys} Set of $1\sigma$ systematic errors
$\{c_n^k\}$ generated by several sources ($k=1,\dots,11$) and
affecting in different ways the binned rates of the SK zenith
distribution $(n=1,\dots,55)$. All errors are given as percentage
fraction of the theoretical rate in each bin: $s_n^k=100\times
c_n^k/R_n^\mathrm{theo}$. The formal identification with the
uncertainties $(\alpha,\delta, \dots,\kappa_u)$ of \cite{Kame} is
given in the 2nd row.
See the text for further details.}\\[-2mm]
\hline\hline\\[.5mm]%======================================================================
bin & event &
$[\cos\theta_z^\mathrm{min},\,\cos\theta_z^\mathrm{max}]$
&$s^1_n$&$s^2_n$&$s^3_n$&$s^4_n$&$s^5_n$&$s^6_n$&$s^7_n$&
$s^8_n$&$s^9_n$&$s^{10}_n$&$s^{11}_n$  \\ %
$n$ & class & range
&$(\alpha)$&$(\delta)$&$(\beta_s)$&$(\beta_m)$&$(\varrho)$&$(\varrho_s)$&$(\varrho_t)$&
$(\eta_s)$&$(\eta_m)$&$(\kappa_f)$&$(\kappa_u)$  \\[.5mm] %
\hline\\[.5mm]%----------------------------------------------------------------------------
%                       &    &   &     &  &    &    &     &     &     &
 1&\sE &$[-1.0,-0.8]$&25&\m12&\m3&    0& 0&   0&   0&\m2.7&    0&\p1.0&    0\\ %
 2&\sE &$[-0.8,-0.6]$&25&\m12&\m3&    0& 0&   0&   0&\m2.1&    0&\p0.5&    0\\ %
 3&\sE &$[-0.6,-0.4]$&25&\m12&\m3&    0& 0&   0&   0&\m1.5&    0&\s0.0&    0\\ %
 4&\sE &$[-0.4,-0.2]$&25&\m12&\m3&    0& 0&   0&   0&\m0.9&    0&\m0.5&    0\\ %
 5&\sE &$[-0.2,-0.0]$&25&\m12&\m3&    0& 0&   0&   0&\m0.3&    0&\m1.0&    0\\ %
 6&\sE &$[+0.0,+0.2]$&25&\m12&\m3&    0& 0&   0&   0&\p0.3&    0&\m1.0&    0\\ %
 7&\sE &$[+0.2,+0.4]$&25&\m12&\m3&    0& 0&   0&   0&\p0.9&    0&\m0.5&    0\\ %
 8&\sE &$[+0.4,+0.6]$&25&\m12&\m3&    0& 0&   0&   0&\p1.5&    0&\s0.0&    0\\ %
 9&\sE &$[+0.6,+0.8]$&25&\m12&\m3&    0& 0&   0&   0&\p2.1&    0&\p0.5&    0\\ %
10&\mE &$[+0.8,+1.0]$&25&\m12&\m3&    0& 0&   0&   0&\p2.7&    0&\p1.0&    0\\[1mm] %
11&\mE &$[-1.0,-0.8]$&25& \m4&  0&\m4.6& 0&   0&   0&    0&\m2.7&\p2.0&    0\\ %
12&\mE &$[-0.8,-0.6]$&25& \m4&  0&\m4.6& 0&   0&   0&    0&\m2.1&\p1.0&    0\\ %
13&\mE &$[-0.6,-0.4]$&25& \m4&  0&\m4.6& 0&   0&   0&    0&\m1.5&\s0.0&    0\\ %
14&\mE &$[-0.4,-0.2]$&25& \m4&  0&\m4.6& 0&   0&   0&    0&\m0.9&\m1.0&    0\\ %
15&\mE &$[-0.2,-0.0]$&25& \m4&  0&\m4.6& 0&   0&   0&    0&\m0.3&\m2.0&    0\\ %
16&\mE &$[+0.0,+0.2]$&25& \m4&  0&\m4.6& 0&   0&   0&    0&\p0.3&\m2.0&    0\\ %
17&\mE &$[+0.2,+0.4]$&25& \m4&  0&\m4.6& 0&   0&   0&    0&\p0.9&\m1.0&    0\\ %
18&\mE &$[+0.4,+0.6]$&25& \m4&  0&\m4.6& 0&   0&   0&    0&\p1.5&\s0.0&    0\\ %
19&\mE &$[+0.6,+0.8]$&25& \m4&  0&\m4.6& 0&   0&   0&    0&\p2.1&\p1.0&    0\\ %
20&\mE &$[+0.8,+1.0]$&25& \m4&  0&\m4.6& 0&   0&   0&    0&\p2.7&\p2.0&    0\\[1mm] %
21&\sM &$[-1.0,-0.8]$&25&\m12& +3&    0& 0&   0&   0&\m2.7&    0&\p1.0&    0\\ %
22&\sM &$[-0.8,-0.6]$&25&\m12& +3&    0& 0&   0&   0&\m2.1&    0&\p0.5&    0\\ %
23&\sM &$[-0.6,-0.4]$&25&\m12& +3&    0& 0&   0&   0&\m1.5&    0&\s0.0&    0\\ %
24&\sM &$[-0.4,-0.2]$&25&\m12& +3&    0& 0&   0&   0&\m0.9&    0&\m0.5&    0\\ %
25&\sM &$[-0.2,-0.0]$&25&\m12& +3&    0& 0&   0&   0&\m0.3&    0&\m1.0&    0\\ %
26&\sM &$[+0.0,+0.2]$&25&\m12& +3&    0& 0&   0&   0&\p0.3&    0&\m1.0&    0\\ %
27&\sM &$[+0.2,+0.4]$&25&\m12& +3&    0& 0&   0&   0&\p0.9&    0&\m0.5&    0\\ %
28&\sM &$[+0.4,+0.6]$&25&\m12& +3&    0& 0&   0&   0&\p1.5&    0&\s0.0&    0\\ %
29&\sM &$[+0.6,+0.8]$&25&\m12& +3&    0& 0&   0&   0&\p2.1&    0&\p0.5&    0\\ %
30&\sM &$[+0.8,+1.0]$&25&\m12& +3&    0& 0&   0&   0&\p2.7&    0&\p1.0&    0\\[1mm] %
31&\mM &$[-1.0,-0.8]$&25& \m4&  0&\p4.6&+6&   0&   0&    0&\m2.7&\p1.0&    0\\ %
32&\mM &$[-0.8,-0.6]$&25& \m4&  0&\p4.6&+6&   0&   0&    0&\m2.1&\p0.5&    0\\ %
33&\mM &$[-0.6,-0.4]$&25& \m4&  0&\p4.6&+6&   0&   0&    0&\m1.5&\s0.0&    0\\ %
34&\mM &$[-0.4,-0.2]$&25& \m4&  0&\p4.6&+6&   0&   0&    0&\m0.9&\m0.5&    0\\ %
35&\mM &$[-0.2,-0.0]$&25& \m4&  0&\p4.6&+6&   0&   0&    0&\m0.3&\m1.0&    0\\ %
36&\mM &$[+0.0,+0.2]$&25& \m4&  0&\p4.6&+6&   0&   0&    0&\p0.3&\m1.0&    0\\ %
37&\mM &$[+0.2,+0.4]$&25& \m4&  0&\p4.6&+6&   0&   0&    0&\p0.9&\m0.5&    0\\ %
38&\mM &$[+0.4,+0.6]$&25& \m4&  0&\p4.6&+6&   0&   0&    0&\p1.5&\s0.0&    0\\ %
39&\mM &$[+0.6,+0.8]$&25& \m4&  0&\p4.6&+6&   0&   0&    0&\p2.1&\p0.5&    0\\ %
40&\mM &$[+0.8,+1.0]$&25& \m4&  0&\p4.6&+6&   0&   0&    0&\p2.7&\p1.0&    0\\[1mm] %
41&\pM &$[-1.0,-0.8]$&25&  +4&  0&    0& 0&+9.2&+5.6&    0&    0&    0&    0\\ %
42&\pM &$[-0.8,-0.6]$&25&  +4&  0&    0& 0&+9.2&+5.6&    0&    0&    0&    0\\ %
43&\pM &$[-0.6,-0.4]$&25&  +4&  0&    0& 0&+9.2&+5.6&    0&    0&    0&    0\\ %
44&\pM &$[-0.4,-0.2]$&25&  +4&  0&    0& 0&+9.2&+5.6&    0&    0&    0&    0\\ %
45&\pM &$[-0.2,-0.0]$&25&  +4&  0&    0& 0&+9.2&+5.6&    0&    0&    0&    0\\[1mm] %
46&\tM &$[-1.0,-0.9]$&25& +12&  0&    0& 0&+9.2&   0&    0&    0&    0&\p1.8\\ %
47&\tM &$[-0.9,-0.8]$&25& +12&  0&    0& 0&+9.2&   0&    0&    0&    0&\p1.4\\ %
48&\tM &$[-0.8,-0.7]$&25& +12&  0&    0& 0&+9.2&   0&    0&    0&    0&\p1.0\\ %
49&\tM &$[-0.7,-0.6]$&25& +12&  0&    0& 0&+9.2&   0&    0&    0&    0&\p0.6\\ %
50&\tM &$[-0.6,-0.5]$&25& +12&  0&    0& 0&+9.2&   0&    0&    0&    0&\p0.2\\ %
51&\tM &$[-0.5,-0.4]$&25& +12&  0&    0& 0&+9.2&   0&    0&    0&    0&\m0.2\\ %
52&\tM &$[-0.4,-0.3]$&25& +12&  0&    0& 0&+9.2&   0&    0&    0&    0&\m0.6\\ %
53&\tM &$[-0.3,-0.2]$&25& +12&  0&    0& 0&+9.2&   0&    0&    0&    0&\m1.0\\ %
54&\tM &$[-0.2,-0.1]$&25& +12&  0&    0& 0&+9.2&   0&    0&    0&    0&\m1.4\\ %
55&\tM &$[-0.1,-0.0]$&25& +12&  0&    0& 0&+9.2&   0&    0&    0&    0&\m1.8\\[1mm] %
\hline\hline%==============================================================================
\end{longtable}
\end{table}
\endgroup
%==================================================================================

Table~\ref{SKsys} shows our numerical assignments for the
$c^k_n$'s (basically derived from \cite{Kame}),  interpreted as
percentage errors of the $R_n^\mathrm{theo}$ values (with or
without oscillations). The second row of Table~\ref{SKsys} reports
the notation of \cite{Kame} for the same error sources, which we
now discuss. The 1st systematic error source ($\alpha$) represents
a 25\% overall normalization uncertainty, due to at least two
components: atmospheric $\nu$ flux normalization error ($\sim\!
20\%$) \cite{Ga03} and $\nu$ cross section uncertainties ($\sim\!
15\%$), added in quadrature \cite{Kiba}. The 2nd error source
($\delta$) is the ``slope'' uncertainty (5\% \cite{Kame}) in the
atmospheric $\nu$ energy spectrum, leading to a $10^{0.05}\simeq
1+12\%$ normalization change for a $1\sigma$ ``tilt'' of the
spectrum slope over one energy decade. Since the SG and UT parent
neutrino spectra are separated by about two decades in energy, we
take for them $c_n^2=-12\%$ and $+12\%$, respectively. The MG and
US event samples have roughly intermediate energies between the SG
and UT ones (on a $\log E_\nu$ scale), so we assign them
$c_n^2=-4\%$ and $+4\%$, respectively. The 3rd and 4th error
sources ($\beta_s$ and $\beta_m$) are the overall $\mu/e$ flavor
ratio uncertainties for the SG and MG samples, respectively. The
relatively large ring-counting error in the MG sample makes
$\beta_m$ greater than $\beta_s$ \cite{Kame}. The 5th error source
($\rho$) is the relative normalization error between partially and
fully contained (PC and FC) events, estimated as $\sim\!10.5\%$
\cite{Kame}. Since only the MG$\mu$ sample contains PC events
($\sim 57\%$), the total (FC+PC) change in the MG$\mu$
normalization is reduced to $0.57 \times 10.5\%\simeq 6\%$, as
indicated in Table~\ref{SKsys}. The 6th error source ($\rho_s$) is
the normalization uncertainty of upgoing muon events (US and UT)
with respect to lower-energy events, mainly due to the relative
cross section uncertainties. The 7th error source ($\rho_t$) gives
further allowance for a normalization shift of the US event rate
only, dominated by the track length cut uncertainty. The 8th and
9th  error sources ($\eta_s$ and $\eta_m$) allow a small up-down
asymmetry (about $\pm3\%$ at $1\sigma$) in the SG and MG zenith
distributions, due to geomagnetic and instrumental uncertainties.
The 10th error source ($\kappa_f$) allows a small
horizontal/vertical ratio uncertainty (about $ \pm 2\%$) in the
zenith distributions of FC events with momentum $>0.4$ GeV for
both the SG and MG samples, mainly related to atmospheric $\nu$
flux calculation uncertainties \cite{Kame}. This error is reduced
by a factor of $\sim 2$ in the SG$e$ and SG$\mu$ samples, due to a
$\sim 1/2$ component of low-momentum FC events which are
unaffected by this uncertainty, since they loose memory of the
neutrino direction. A similar $\kappa_f$ error reduction applies
to the MG$\mu$ sample, due to the substantial presence of
(unaffected) PC events \cite{Kame}. The 11th error source
($\kappa_u$) gives allowance for a horizontal/vertical ratio
uncertainty in the UT$\mu$ sample, which is instead argued to be
irrelevant in the US$\mu$ sample (see \cite{Kame}). Finally, we
remark that the SK analysis contains a further error source
(so-called $L/E_\nu$ uncertainty, $\sim\!15\%$
\cite{Kame,Kiba,Ishi,Mess}) that we {\em do not\/} implement in
our analysis. In particular, we think that the spread in $L$
should be more properly included through integration over the {\em
known\/} neutrino pathlength distribution (i.e., over the neutrino
production heights at different angles) rather
than being interpreted as an additional uncertainty.%
%----------------------------------------------------------------
\footnote{We understand that future SK atmospheric $\nu$ data
analyses might indeed fully incorporate the spread of $L$, as
derived from recent three-dimensional $\nu$ flux calculations
(T.~Kajita, private communication).}
%--------------------------------------------------------------------
 We also observe that, in general, a ``$L/E_\nu$ error'' becomes
meaningless in scenarios where $P_{\mu\mu}$ is not a function of
$L/E_\nu$, but of $L$ and $E_\nu$ separately (as is the case,
e.g., in the presence of matter effects).

The 11 systematic error sources listed in Table~\ref{SKsys} are
not necessarily independent. For instance, the horizontal/vertical
ratio uncertainties $\eta_s$ and $\eta_m$ are largely due to
common geomagnetic and instrumental effects, and are thus expected
to be positively correlated. On the other hand, cross section
uncertainties (affecting in different ways quasi-elastic and
deep-inelastic scattering events) can generate a ``migration'' of
events from one class to another, thus inducing negative
correlations among the corresponding normalization uncertainties.

%==================================================================================
\begingroup\squeezetable
\begin{table}[t]
\newcommand{\m}{$-$}
\newcommand{\p}{$+$}
\newcommand{\s}{\phantom{$+$}}
\caption{\label{SKcor} \footnotesize\baselineskip=6pt Correlation
matrix $\rho_{hk}$ of systematic error sources $(h,k=1,\dots,11)$
in the SK analysis, as taken from \cite{Kame}.}
\begin{ruledtabular}
\begin{tabular}{r|ccccccccccc}
%===========================================================================================
$\rho$ & 1 & 2 & 3 & 4 & 5 & 6 & 7 & 8 & 9 & 10 & 11  \\[.5mm] %
\hline%-------------------------------------------------------------------------------------
1   &\p1.00&   0  &   0  &   0  &   0  &   0  &   0  &   0  &   0  &   0  &   0  \\ %
2   &      &\p1.00&   0  &   0  &   0  &   0  &   0  &   0  &   0  &   0  &   0  \\ %
3   &      &      &\p1.00&\p0.31&\m0.17&\m0.03&   0  &   0  &   0  &   0  &   0  \\ %
4   &      &      &      &\p1.00&\m0.34&   0  &   0  &   0  &   0  &   0  &   0  \\ %
5   &      &      &      &      &\p1.00&\p0.58&\m0.17&   0  &   0  &   0  &   0  \\ %
6   &      &      &      &      &      &\p1.00&\m0.25&   0  &   0  &   0  &   0  \\ %
7   &      &      &      &      &      &      &\p1.00&   0  &   0  &   0  &   0  \\ %
8   &      &      &      &      &      &      &      &\p1.00&\p0.62&   0  &   0  \\ %
9   &      &      &      &      &      &      &      &      &\p1.00&   0  &   0  \\ %
10  &      &      &      &      &      &      &      &      &      &\p1.00&   0  \\ %
11  &      &      &      &      &      &      &      &      &      &      &\p1.00\\ %
\end{tabular}
\end{ruledtabular}
\end{table}
\endgroup
%===========================================================================================
Table~\ref{SKcor} reports explicitly the correlation matrix
$\rho_{hk}$ among the SK systematics, as taken from the careful
estimates in \cite{Kame}. This matrix enters in the
$\chi^2_\mathrm{SK}$ definition of Eq.~(\ref{chiSK}).

Finally, Table~\ref{SKpull} reports the systematic SK pulls
$\{\overline\xi_k\}_{k=1,\dots,11}$ at the global (SK+K2K) best
fit for standard $\nu_\mu\to\nu_\tau$ oscillations, which have
been used to calculate the ``shifted'' theoretical rates in
Fig.~\ref{fig04}. They appear to be relatively small for any $k$.
Therefore, the induced differences $\overline
R_n^\mathrm{theo}-R_n^\mathrm{theo}$ between the solid and dashed
histograms in Fig.~\ref{fig04} represent tolerable systematic
shifts of ``typical'' size.

%%%%%%%%%%%%%%%%%%%%%%%%%%%%%%%%%%%%%%%%%%%%%%%%%%%%%%%%%%%%%%%%%%%%%%%%%
\begingroup\squeezetable
\begin{table}[t]
\caption{\label{SKpull} \footnotesize\baselineskip=6pt Values of
the SK systematic pulls $\{\overline\xi_k\}$ at the global
(SK+K2K) best-fit point for standard $\nu_\mu\to\nu_\tau$
oscillations.}
\newcommand{\m}{$-$}
\begin{ruledtabular}
\begin{tabular}{cccccccccccc}
$k$ & 1 & 2 & 3 & 4 & 5 & 6 & 7 & 8 & 9 & 10 & 11 \\ %
$\overline\xi_k$ & +0.19 & +0.22 & \m1.05 & \m0.99 & \m0.53 &
\m0.03 & \m0.12 & \m0.99 &
\m0.82 & +0.07 & +0.01 \\
\end{tabular}
\end{ruledtabular}
\end{table}
\endgroup

%%%%%%%%%%%%%%%%%%%%%%%%%%%%%%%%%%%%%%%%%%%%%%%%%%%%%%%%%%%%%%%%%%%%%%%%%%%%%%%%%%%%%%%
\section{Calculation and statistical analysis of the K2K spectrum}

In this appendix we describe our calculation of the K2K spectrum
of events and our approach to the K2K statistical analysis.

\subsection{Calculation of the K2K spectrum}

Our oscillation analysis of the K2K spectral data is based on 29
single-ring $\mu$-like (1R$\mu$) events \cite{K2K2}, binned into 6
intervals of the reconstructed neutrino energy $E$. The
experimental number of events $N_n^\mathrm{expt}$ in each bin is
given in Table~\ref{K2Ksys}, together with the corresponding
systematic uncertainties (discussed below). The no-oscillation
expectations correspond to a total of 42.5 events, as graphically
derived from Fig.~2 in \cite{K2K2,Wi02}. This data sample contains
mainly muons
from quasi-elastic (QE) scattering $\nu_\mu+n\to\mu+p$.%
%----------------------------------------------------------------
\footnote{We do not use the additional 27 events forming the total
K2K data sample (56 events), since the simulation of (dominantly)
non-QE events is not possible without very detailed K2K
experimental information.}
%---------------------------------------------------------------

%==================================================================================
\begingroup\squeezetable
\begin{table}[t]
\newcommand{\m}{$-$}
\newcommand{\p}{$+$}
\newcommand{\s}{\phantom{$+$}}
\caption{\label{K2Ksys} \footnotesize\baselineskip=6pt Set of
systematic errors $\{c_n^k\}$ generated by several sources
$(k=1,\dots,7)$, and affecting in different ways the binned rates
of the K2K energy distribution $(n=1,\dots,6)$. All errors are
given as  percentage fraction of the theoretical number of events
$N_n^\mathrm{theo}$ in each bin: $s_n^k=100\times
c_n^k/N_n^\mathrm{theo}$. The energy interval and the experimental
number of events $N_n^\mathrm{expt}$ are given in the second and
third row, respectively. See the text for details.}
\begin{ruledtabular}
\begin{tabular}{c|ccccccccc}
%===========================================================================================
$n$ & $E$ range (GeV) & $N_n^\mathrm{expt}$ &$s^1_n$&$s^2_n$&$s^3_n$&$s^4_n$&$s^5_n$&$s^6_n$&$s^7_n$  \\[.5mm] %
\hline%-------------------------------------------------------------------------------------
1 & $[0.0,0.5]$ & 3 & 2.6&   0&   0&    0&    0&    0& 5.0\\%
2 & $[0.5,1.0]$ & 4 &   0& 4.3&   0&    0&    0&    0& 5.0\\%
3 & $[1.0,1.5]$ &14 &   0&   0& 6.5&    0&    0&    0& 5.0\\%
4 & $[1.5,2.0]$ & 2 &   0&   0&   0& 10.4&    0&    0& 5.0\\%
5 & $[2.0,2.5]$ & 4 &   0&   0&   0&    0& 11.1&    0& 5.0\\%
6 & $>2.5$      & 2 &   0&   0&   0&    0&    0& 12.2& 5.0\\%
\end{tabular}
\end{ruledtabular}
\end{table}
\endgroup
%===========================================================================================

In principle, for QE scattering on neutrons at rest, the
reconstructed neutrino energy $E$ depends only on the muon
kinematics \cite{K2K2},
%...................................................................
\begin{equation}
\label{Erec}
 E=\frac{m_n E_\mu -
m^2_\mu/2}{m_n-E_\mu+p_\mu\cos\theta_\mu}\ .
\end{equation}
%...................................................................
However, a number of effects can produce significant deviations of
$E$ from the {\em true\/} neutrino energy $E_\nu$, including
nuclear effects (Fermi motion, Pauli blocking, proton
re-scattering \cite{Ca01,Wa02,Wa03}), detection uncertainties (SK
resolution in energy and angle \cite{Ca02,Ca00}), and
contamination from non-QE events \cite{It02,It03}. Therefore, the
reconstructed $\nu$ energy $E$ is distributed around the true
$\nu$ energy $E_\nu$ through a certain (normalized) probability
distribution function $R(E,E_\nu)$. This function smears out the
oscillation pattern and must thus be taken into account, at least
approximately, in the K2K analysis.

Although the function $R(E,E_\nu)$ is not explicitly provided by
the K2K Collaboration \cite{K2K2}, its main features can be
roughly recovered from other studies. In particular, $R(E,E_\nu)$
has been derived (for a SK-like detector) through numerical
simulations in \cite{Ca01,Ca02}. It turns out that
\cite{Ca01,Ca02}: 1) $E$ is typically biased towards slightly
lower values than $E_\nu$ (with $\langle E_\nu - E\rangle = b
\simeq
\mathrm{few}\times 10$~MeV),%
%------------------------------------------------------------------
\footnote{Contamination from non-QE events also produces a bias
$b$ in the same direction \cite{It02}.}
%-------------------------------------------------------------------
and 2) the r.m.s.\ difference $\langle
E-E_\nu\rangle_\mathrm{rms}=\sigma_E$ increases from zero (at
$E\simeq 0.2$~MeV) up to a plateau $\sigma_E/E\simeq 20\%$
\cite{Ca01,Ca02}. We approximately embed these features in a
Gaussian p.d.f.\ of the kind
%.....................................................................................
\begin{equation}
\label{R} R(E,E_\nu)\simeq
\frac{1}{\sqrt{2\pi}\sigma_E}e^{-\frac{1}{2}\left(\frac{E-E_\nu+b}{\sigma_E}\right)^2}\
,
\end{equation}
%.....................................................................................
where we take $b\simeq 0.05$~GeV for the bias parameter and
%.....................................................................................
\begin{equation}
\label{sigmaE} \sigma_E/E_\nu\simeq
0.2\left[1-\exp\left(\frac{0.2-E_\nu[\mathrm{GeV}]}{0.08}\right)\right]
\end{equation}
%.....................................................................................
for the energy reconstruction error $\sigma_E$ above
$E_\nu>0.2$~GeV ($\sigma_E=0$ otherwise). Although we do not
include the more detailed features investigated in
\cite{Ca01,Ca02} (e.g., asymmetric tails), the above function
$R(E,E_\nu)$ appears to be good enough for our current analysis. A
better treatment will be possible when a description of
$R(E,E_\nu)$ (or of equivalent
information) in K2K will be made publicly available.%
%----------------------------------------------------------------------------------
\footnote{We also note that several issues related to the
characterization of the $E-E_\nu$ difference in QE and non-QE
events are currently being re-examined, in the light of the
accuracy needed to analyze higher-statistics data in the K2K and
future low-energy accelerator neutrino experiments (see, e.g.,
\cite{Ca03,Wa03,It03}).}
%-----------------------------------------------------------------------------------

The $E$-spectrum of events expected in the presence of
oscillations is calculated as
%.....................................................................................
\begin{equation}
\label{dNdE} \frac{dN^\mathrm{theo}}{dE}=\int dE_\nu\, S(E_\nu)\,
R(E,E_\nu) \, P_{\mu\mu}(E_\nu)\ ,
\end{equation}
%.....................................................................................
where $S(E_\nu)$ is the unoscillated $\nu$ spectrum at the SK
detector (in terms of the true neutrino energy).  We find that, in
the relevant energy range $E_\nu\simeq[0.2,3.25]$~GeV, the K2K
spectrum $S(E_\nu)$ is well approximated by the simple function
%......................................................................................
\begin{equation}
S(E_\nu)\simeq \gamma (E_\nu-0.2)^\alpha(3.25-E_\nu)^\beta\ ,
\end{equation}
%......................................................................................
where $\alpha=1.0056$, $\beta=3.144$, and $E_\nu$ is given in GeV.
The factor $\gamma=0.7389$ is then fixed by the normalization
constraint $\int dE_\nu\, dE\, S(E_\nu)\, R(E,E_\nu)=42.5$ for no
oscillation. The reader is referred to Fig.~\ref{fig09} for
representative theoretical spectra $dN^\mathrm{theo}/dE$, with and
without oscillations (at maximal mixing). Such spectra are in
reasonable agreement with the corresponding K2K ones as shown in
\cite{Oy02}.

\subsection{Statistical analysis of the K2K spectrum}

We bin both the K2K theoretical spectrum $dN^\mathrm{theo}/dE$ and
the experimental spectrum into six energy intervals (see
Table~\ref{K2Ksys}). The statistical analysis of the K2K absolute
number of events $\{N_n^X\}_{n=1,\dots,6}$
($X=\mathrm{theo},\,\mathrm{expt}$) is performed, in analogy with
the previous Appendix, by shifting the  predictions through a set
of (7, in this case) systematic deviations
%..........................................................................
\begin{equation}
\label{Ntilde} N_n^\mathrm{theo}\to \tilde
N_n^\mathrm{theo}=N_n^\mathrm{theo}+\sum_{k=1}^{7}\xi_k c_n^k\ ,
\end{equation}
%.........................................................................
which are then minimized away in the $\chi^2$. The only formal
difference with respect to the SK data fit is the use of Poisson
statistics, required to deal with small number of events:
%..........................................................................
\begin{equation}
\label{chiK2K}\chi^2_\mathrm{K2K}=\min_{\{\xi_k\}} %
\left[ 2\sum_{n=1}^6\left(\tilde
N_n^\mathrm{theo}-N_n^\mathrm{expt} - N_n^\mathrm{exp}\ln
\frac{\tilde N_n^\mathrm{theo}}{N_n^\mathrm{expt}}\right)+
\sum_{k,h=1}^7 \xi_k \,[\rho^{-1}]_{hk} \,\xi_h \right]\ .
\end{equation}
%.........................................................................
The minimization in Eq.~(\ref{chiK2K}), which can be easily
performed through linearization in the small parameters $\xi_k$,
fixes the best-fit systematic pulls $\overline\xi_k$. The shifted
theoretical predictions for the number of events in each bin read
then
%..........................................................................
\begin{equation}
\label{Nbar} \overline
N_n^\mathrm{theo}=N_n^\mathrm{theo}+\sum_{k=1}^{7}\overline \xi_k
c_n^k\ .
\end{equation}
%.........................................................................

In the systematic error budget, we include the following
(dominant) error sources, for which one can find a detailed
description in \cite{Ko02}. The near-far extrapolation procedure
in K2K leads to six systematics errors $(k=1,\dots,6)$ in each
bin, with significant bin-to-bin correlations \cite{Ko02}. The
overall normalization uncertainty adds a further 5\% error ($k=7$)
with full correlation in each bin. The numerical values of the
$1\sigma$ systematics $c^n_k$ and of the correlation matrix
$\rho_{hk}$, as derived from \cite{Ko02}, are reported in
Tables~\ref{K2Ksys} and \ref{K2Kcor}, respectively. These values
enter in the $\chi^2_\mathrm{K2K}$ definition of
Eq.~(\ref{chiK2K}).

\begin{figure}
\vspace*{-0.2cm}\hspace*{-1.2cm}
\includegraphics[scale=0.92, bb= 30 100 500 700]{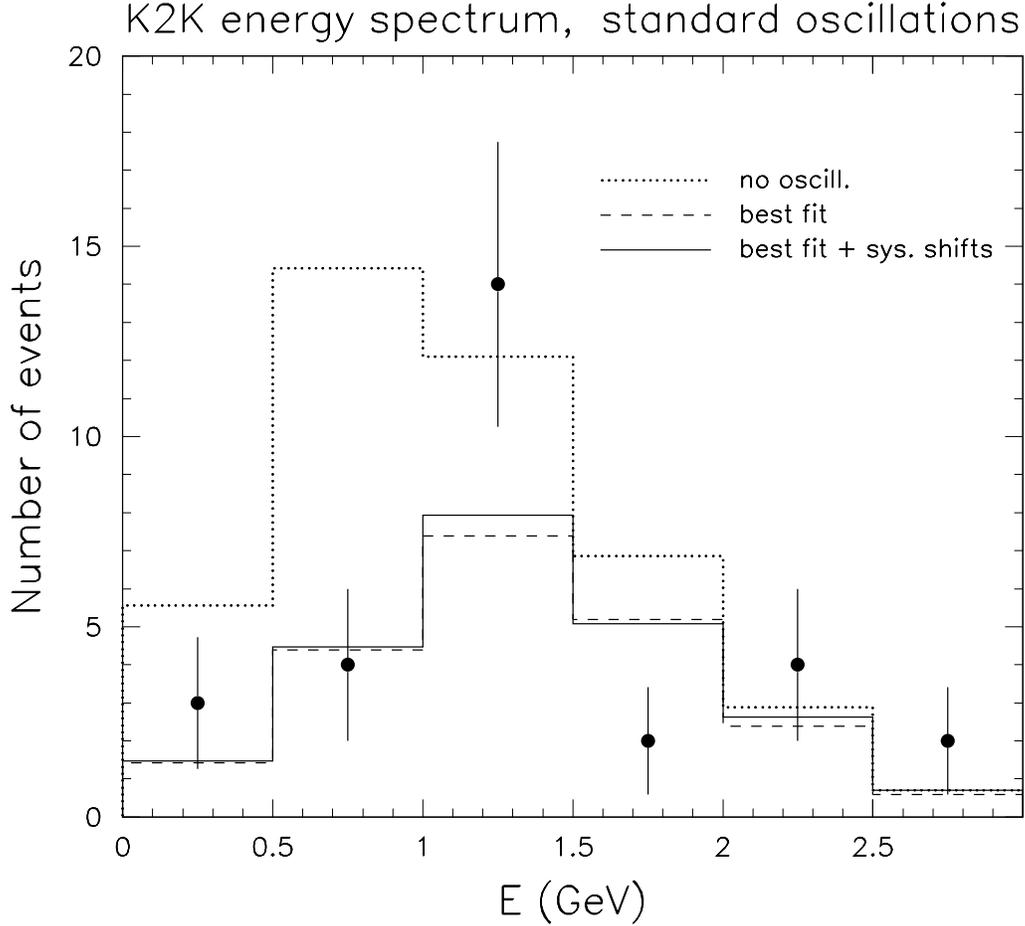}
\vspace*{-2cm} \caption{\label{fig05} Standard oscillations in the
$\nu_\mu\to\nu_\tau$ channel: Absolute spectrum of (dominantly QE)
events in K2K, as a function of the reconstructed neutrino energy
$E$. The data points (29 events total) are shown as dots with
$\pm1\sigma_\mathrm{stat}$ in each of the six bins. The dotted
histogram represents our calculations for no oscillation. The
solid and dashed histograms represent the theoretical predictions
$N_{n}^\mathrm{theo}$ and $N_{n}^\mathrm{theo}$ at the global
(SK+K2K) best-fit point in Table~\ref{chi2val}, with and without
systematic shifts, respectively. See the text and Appendix~B  for
details.}
\end{figure}

\begin{figure}
\vspace*{-2.5cm}\hspace*{-1.8cm}
\includegraphics[scale=0.92, bb= 30 100 500 700]{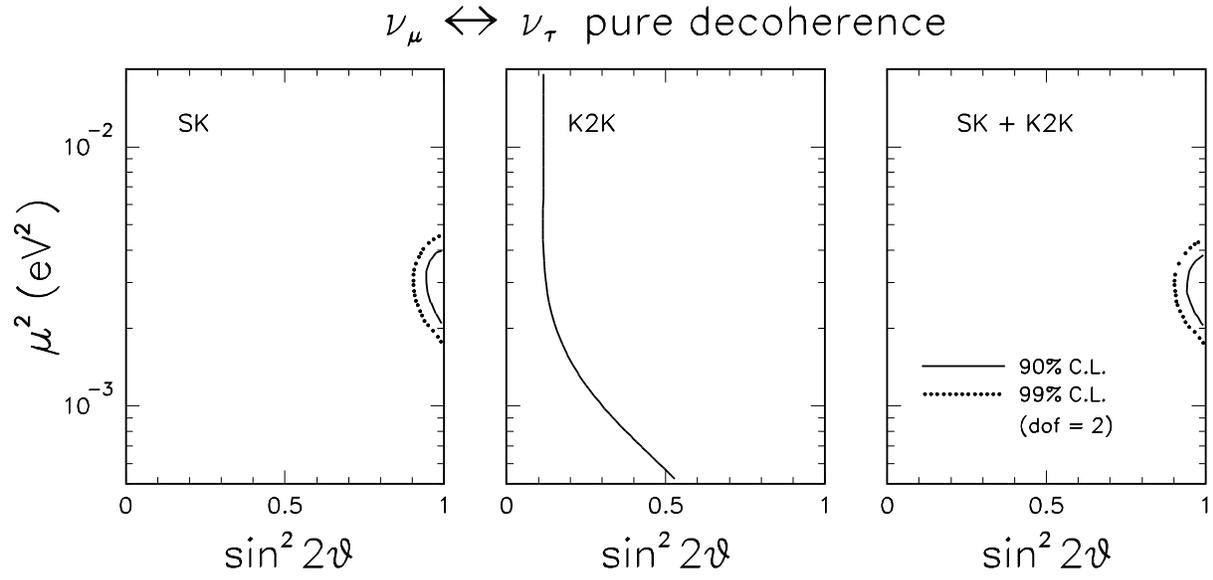}
\vspace*{-2cm} \caption{\label{fig06} As in Fig.~\ref{fig01}, but
for the pure decoherence scenario. Notice that the bounds on the
$(\mu^2,\,\sin^22\theta)$ parameters are dominated by SK.}
\end{figure}

\begin{figure}
\vspace*{-1.5cm}\hspace*{-3cm}
\includegraphics[scale=0.92, bb= 30 100 500 700]{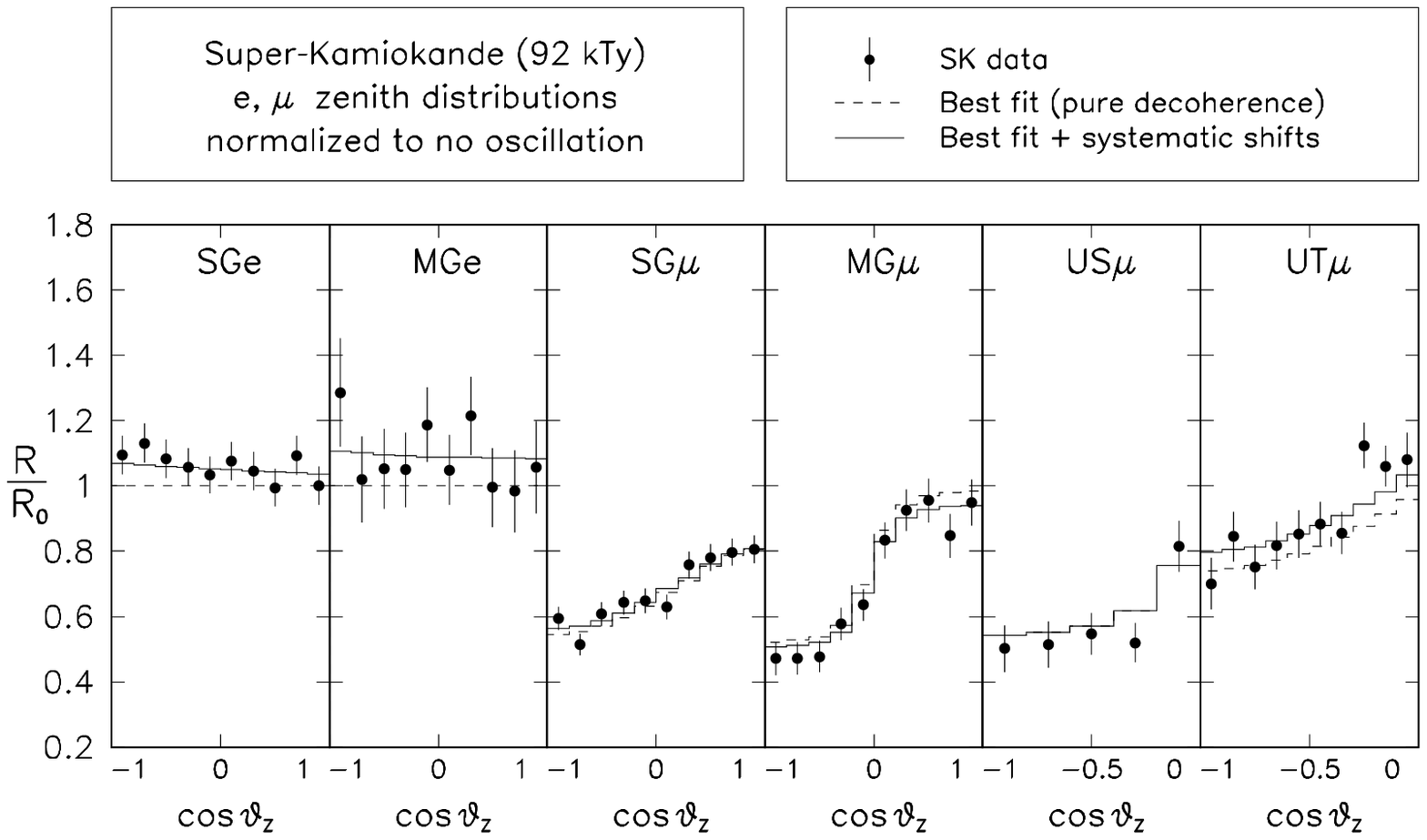}
\vspace*{-3cm} \caption{\label{fig07} As in Fig.~\ref{fig04}, but
for the global (SK+K2K) best-fit point to the pure decoherence
scenario, as given in Table~\ref{chi2val}.}
\end{figure}

\begin{figure}
\vspace*{-0.2cm}\hspace*{-1.2cm}
\includegraphics[scale=0.92, bb= 30 100 500 700]{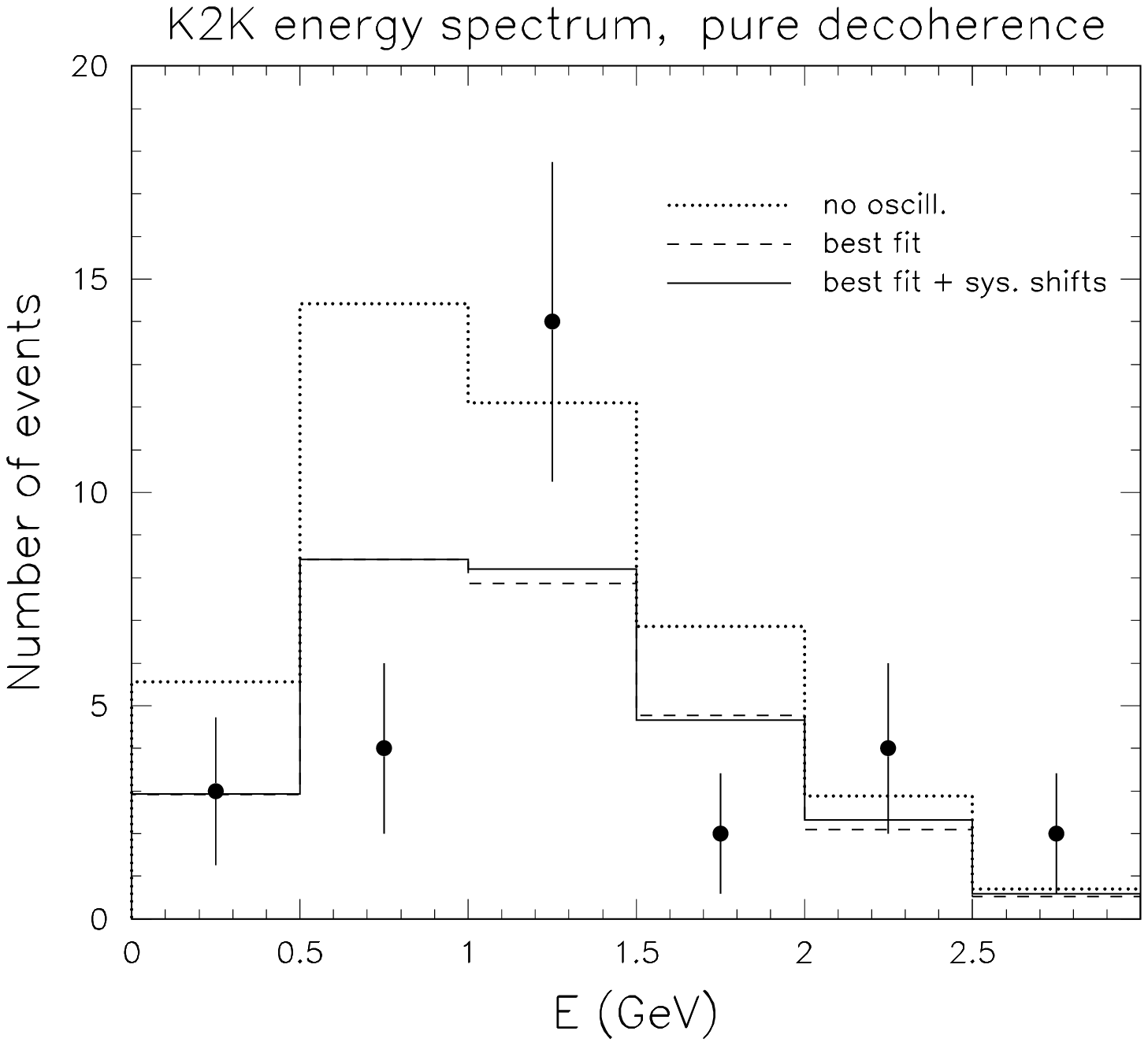}
\vspace*{-2cm} \caption{\label{fig08} As in Fig.~\ref{fig05}, but
for the global (SK+K2K) best-fit point to the pure decoherence
scenario, as given in Table~\ref{chi2val}.}
\end{figure}

\begin{figure}
\vspace*{-0.2cm}\hspace*{-1.2cm}
\includegraphics[scale=0.92, bb= 30 100 500 700]{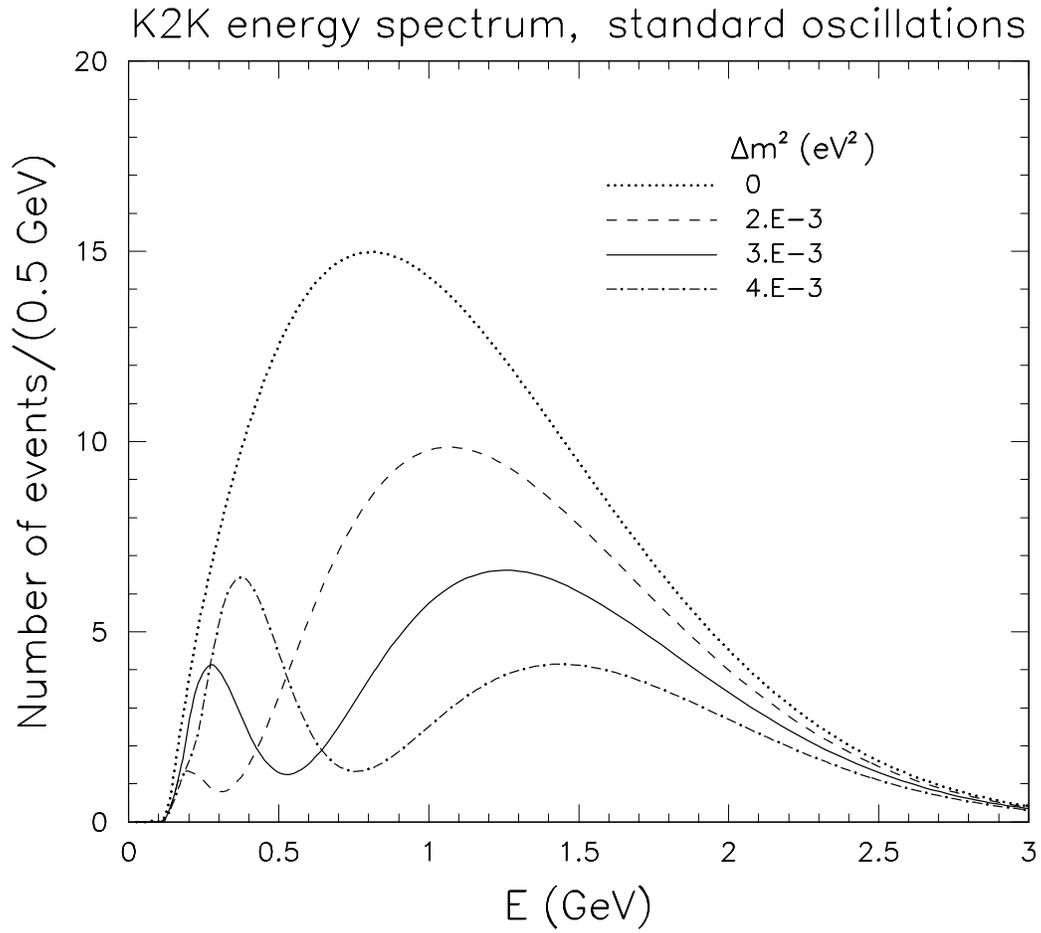}
\vspace*{-2cm} \caption{\label{fig09} Standard oscillations:
Unbinned K2K theoretical spectrum at maximal mixing and for three
representative values of $\Delta m^2$.}
\end{figure}

\begin{figure}
\vspace*{-0.2cm}\hspace*{-1.2cm}
\includegraphics[scale=0.92, bb= 30 100 500 700]{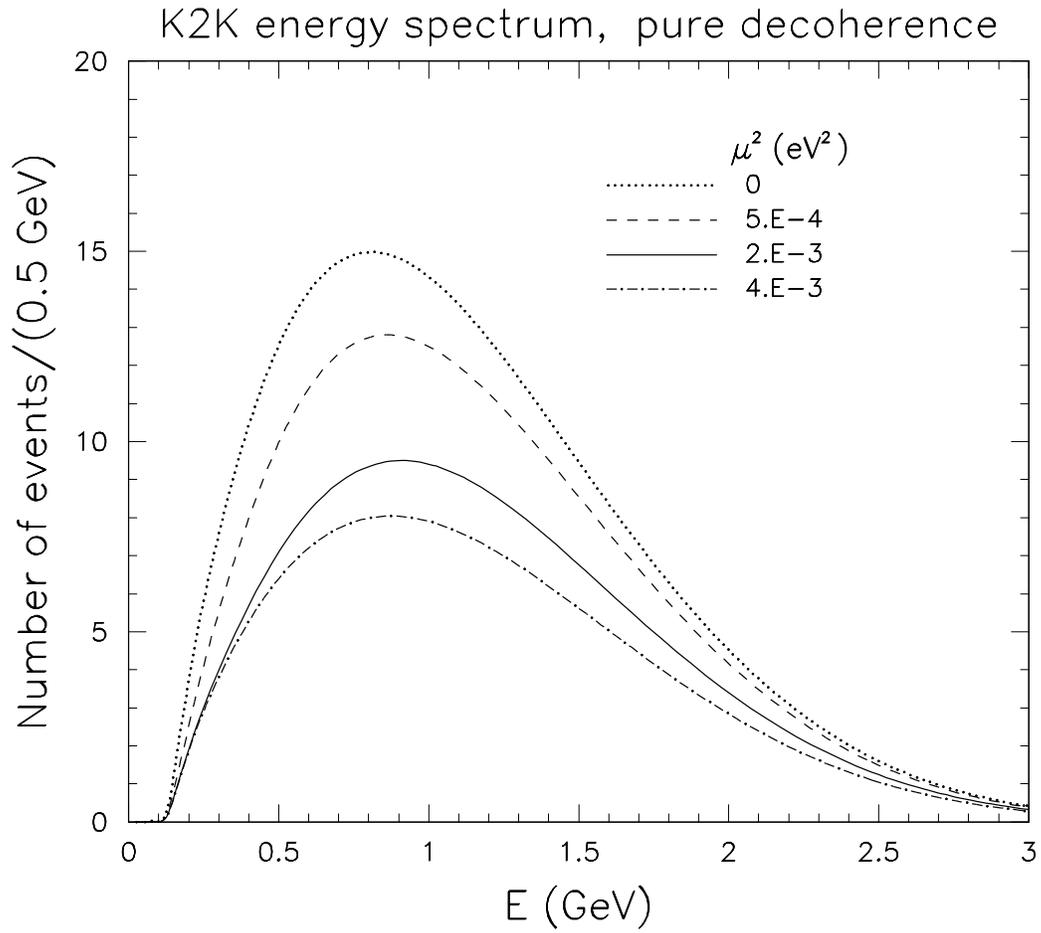}
\vspace*{-2cm} \caption{\label{fig10} Pure decoherence: Unbinned
K2K theoretical spectrum at maximal mixing and for three
representative values of $\mu^2$.}
\end{figure}

\begin{figure}
\vspace*{+1.cm}\hspace*{-2.1cm}
\includegraphics[scale=0.92, bb= 30 100 500 700]{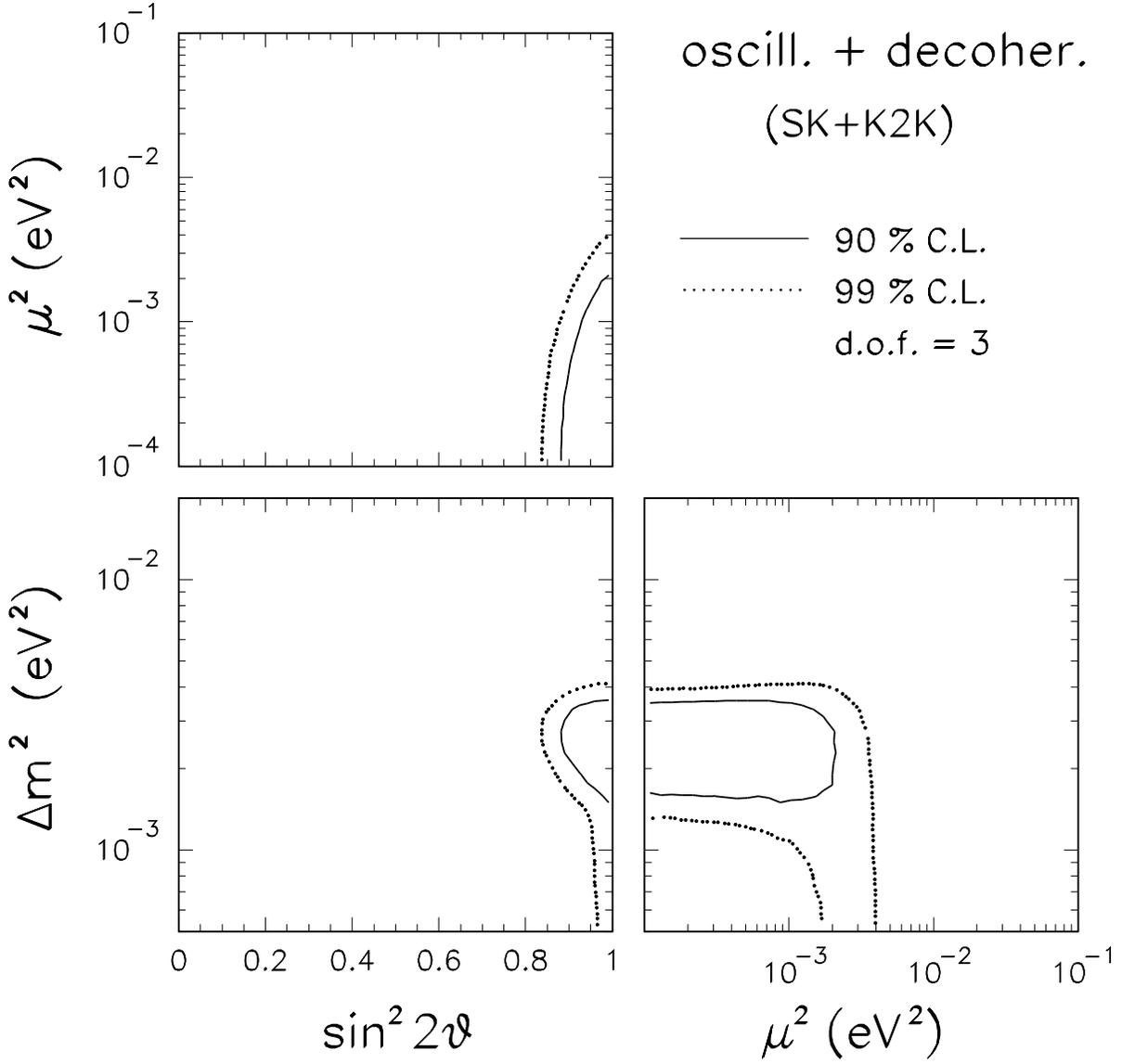}
\vspace*{-2cm} \caption{\label{fig11} General case (oscillations
plus decoherence): Volume allowed by SK+K2K in the parameter space
$(\Delta m^2,\sin^2 2\theta, \mu^2)$  at 90\% and 99\% C.L.\ for
$N_\mathrm{DF}=3$ ($\Delta \chi^2=6.25$ and 11.34), shown through
its projections onto the coordinate planes. The best-fit is
reached in the limit of standard oscillations ($\mu^2=0$).
However, the case of pure decoherence ($\Delta m^2=0$) is still
marginally allowed. In all cases, the mixing is nearly maximal
$(\sin^2 2\theta\simeq 1)$ }
\end{figure}

\begin{figure}
\vspace*{+2cm}\hspace*{-2.6cm}
\includegraphics[scale=0.92, bb= 30 100 500 700]{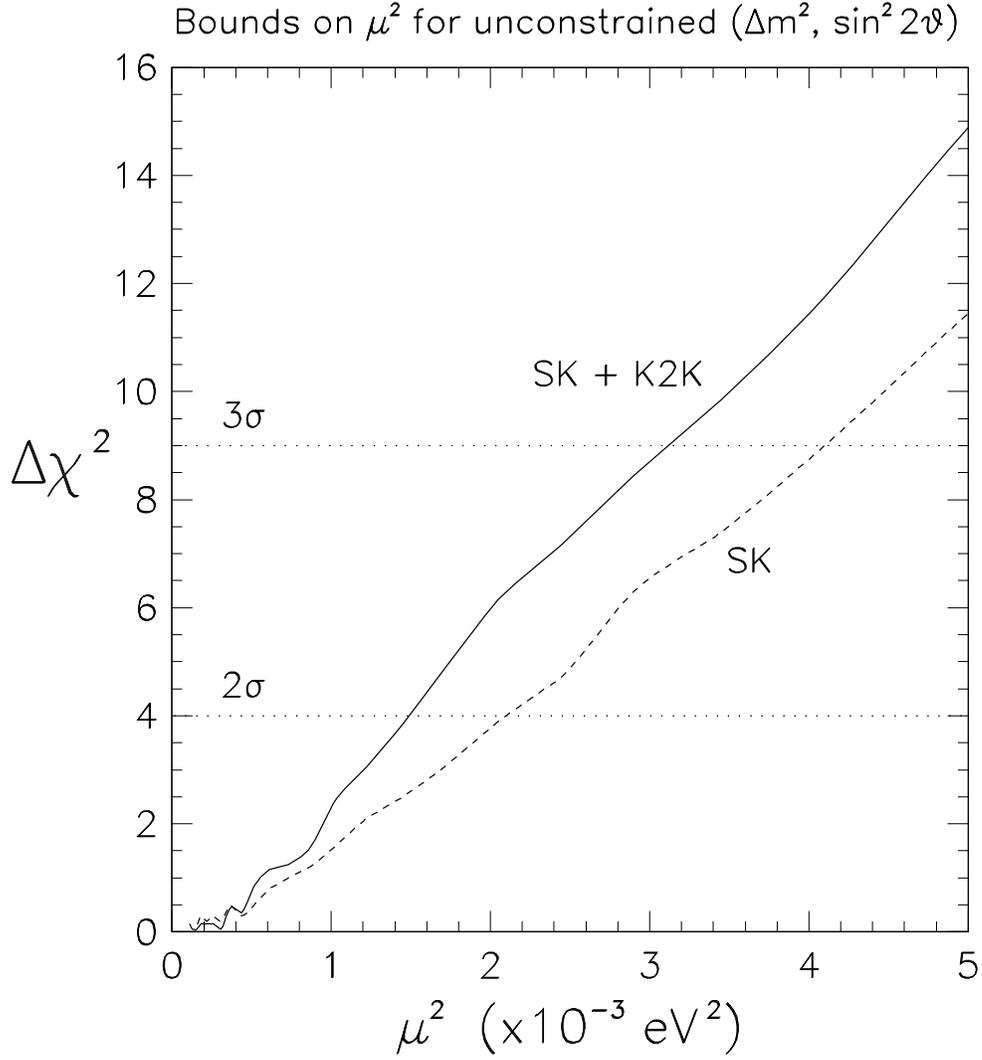}
\vspace*{-2cm} \caption{\label{fig12} General case (oscillations
plus decoherence): Upper bounds on the decoherence parameter
$\mu^2$, for unconstrained $(\delta m^2,\,\sin^2 2\theta)$. The
intersections with the horizontal dotted lines give the $2\sigma$
and $3\sigma$ upper bounds on $\mu^2$ for $N_\mathrm{DF}=1$.}
\end{figure}

Finally, Table~\ref{K2Kpull} reports the K2K systematic pulls
$\{\overline\xi_k\}_{k=1,\dots,7}$ at the global (SK+K2K) best fit
for standard $\nu_\mu\to\nu_\tau$ oscillations, from which the
``shifted'' theoretical rates in Fig.~\ref{fig05} are calculated.
All the pulls appear to be relatively small.

\begingroup\squeezetable
%==================================================================================
\begin{table}[t]
\newcommand{\m}{$-$}
\newcommand{\p}{$+$}
\newcommand{\s}{\phantom{$+$}}
\caption{\label{K2Kcor} \footnotesize\baselineskip=6pt Correlation
matrix $\rho_{hk}$ ($h,k=1,\dots,7$) of systematics in the K2K
analysis. The nontrivial $6\times 6$ block, related to near-far
extrapolation uncertainties, is derived from \cite{Ko02}. The 7th
systematic error (independent overall normalization) is assumed to
be uncorrelated.}
\begin{ruledtabular}
\begin{tabular}{r|ccccccc}
%===========================================================================================
$\rho$ & 1 & 2 & 3 & 4 & 5 & 6 & 7   \\[.5mm] %
\hline%-------------------------------------------------------------------------------------
1   &\p1.00&\m0.25&   0  &   0  &   0  &   0  &   0    \\ %
2   &      &\p1.00&   0  &   0  &   0  &   0  &   0    \\ %
3   &      &      &\p1.00&\p0.08&\m0.04&\m0.20&   0    \\ %
4   &      &      &      &\p1.00&\p0.79&\p0.19&   0    \\ %
5   &      &      &      &      &\p1.00&\p0.39&   0    \\ %
6   &      &      &      &      &      &\p1.00&   0    \\ %
7   &      &      &      &      &      &      &\p1.00  \\ %
\end{tabular}
\end{ruledtabular}
\end{table}
%===========================================================================================
\endgroup

\begingroup\squeezetable
\begin{table}[t]
\newcommand{\m}{$-$}
\caption{\label{K2Kpull} \footnotesize\baselineskip=6pt Values of
the K2K systematic pulls $\{\overline\xi_k\}$ at the global
(SK+K2K) best-fit point for standard $\nu_\mu\to\nu_\tau$
oscillations.}
\begin{ruledtabular}
\begin{tabular}{cccccccc}
$k$ & 1 & 2 & 3 & 4 & 5 & 6 & 7  \\ %
$\overline\xi_k$ & +0.02 & \m0.01 & +0.29 & \m0.70 & +0.38 & +0.24 & +0.38 \\
\end{tabular}
\end{ruledtabular}
\end{table}
\endgroup

%%%%%%%%%%%%%%%%%%%%%%%%%%%%%%%%%%%%%%%%%%%%%%%%%%%%%%%%%%%%%%%%%%%%%%%%%%%%%%%%%%%%

\newpage
%%%%%%%%%%%%%%%%%%%%%%%%%%%%%%%%%%%%%%%%%%%%%%%%%%%%%%%%%%%%%%%%%%%%%%%%%%%%%%%%%%%
%---------------------------------------------------------------------------
%---------------------------------------------------------------------------
%---------------------------------------------------------------------------
%---------------------------------------------------------------------------
%---------------------------------------------------------------------------
%---------------------------------------------------------------------------
%---------------------------------------------------------------------------
%---------------------------------------------------------------------------
%---------------------------------------------------------------------------
%---------------------------------------------------------------------------
%---------------------------------------------------------------------------
%---------------------------------------------------------------------------

\end{document}